\documentclass[12pt,a4paper,oneside]{article}
\usepackage[margin=1in]{geometry}
\usepackage{graphicx} 
\usepackage{amsmath,amsthm,amssymb,mathtools}
\usepackage{booktabs}
\usepackage{xcolor}
\usepackage{cleveref}
\usepackage{natbib}
\usepackage{enumitem}

\usepackage{tikz}
\usetikzlibrary{automata,positioning,fit, shapes.geometric,calc}
\usepackage{caption}

\newtheorem{theorem}{Theorem}
\newtheorem{lemma}{Lemma}
\newtheorem{assumption}{Assumption}

\newtheorem{proposition}{Proposition}

\title{Policy-relevant causal effect estimation using instrumental variables with interference}
\author{Didier Nibbering\thanks{Department of Econometrics and Business Statistics, Monash University, Melbourne, Australia. \mbox{didier.nibbering@monash.edu}} \and
Matthijs Oosterveen\thanks{Department of Economics, Lisbon School of Economics and Management, and Advance/ISEG Research, University of Lisbon, Lisbon, Portugal. \mbox{oosterveen@iseg.ulisboa.pt}} 
}
\date{\today}

\begin{document}
\linespread{1.70}
\normalsize
\maketitle
\thispagestyle{empty}

\begin{abstract} 
\noindent {\normalsize 
Many policy evaluations using instrumental variable (IV) methods include individuals who interact with each other, potentially violating the standard IV assumptions. This paper defines and partially identifies direct and spillover effects with a clear policy-relevant interpretation under relatively mild assumptions on interference. Our framework accommodates both spillovers from the instrument to treatment and from treatment to outcomes and allows for multiple peers. By generalizing monotone treatment response and selection assumptions, we derive informative bounds on policy-relevant effects without restricting the type or direction of interference. The results extend IV estimation to more realistic social contexts, informing program evaluation and treatment scaling when interference is present.
} 


\end{abstract}

\newpage
\setcounter{page}{1}

\section{Introduction}
Within the instrumental variable (IV) framework, the outcome of any individual depends only on their own treatment, and not on the treatment  of others. In practice, individuals naturally interact with each other, and this assumption is likely to be violated. This implies that the treatment of one individual may affect the outcome of another individual, which we refer to as treatment interference. For instance, the vaccination status of one individual may reduce the infection risk for others. The potential outcome of an individual now depends not just on its own treatment, but on others' treatments too. This complicates the IV estimation of treatment effects as (1) the standard IV exclusion restriction may be violated which results in biased estimates, and (2) many different treatment effects can be defined.

This paper partially identifies local average direct and spillover effects. Understanding both direct and spillover effects is essential for program evaluation in settings with interference. For example, policymakers may worry that the spillover effects of a job training program offset the direct effects: job placement could merely change who is employed without affecting the overall employment rate. We differentiate the direct effects to individuals with treated and untreated peers. Similarly, we identify spillover effects for treated and untreated individuals. These four effects allow policymakers to anticipate treatment scaling and coverage effects. For instance, if direct and spillovers effects are higher with treated peers, increased or concentrated treatment strategies maximize impacts. In education interventions, for instance, policymakers may decide to target entire classrooms, instead of students across different classes.

The direct effects have a clear interpretation as a local average treatment effect (LATE) of receiving your own treatment while keeping peers' treatment fixed, which provides an advantage over other causal parameters with interference. For instance, \citet{imai2021causal} and \citet{hoshino2024causal} identify weighted averages including LATEs of own treatment and peers' treatment. \citet{kang2016peer} identify effects of own treatment conditional on a certain share of peers assigned to treatment, instead of receiving treatment. Our spillover effects have similar improvements in interpretability, especially when each individual has one peer. In the absence of interference, the direct effects equal the LATE of \citet{imbens1994identification} and the spillover effects equal zero. 

To identify the direct and spillover effects, additional assumptions beyond the standard IV framework are required. First, to limit the number of potential outcomes, we impose the partial interference assumption that each individual only has interactions with a small number of known peers. Second, we extend the standard IV monotonicity assumption to the spillovers: an individual is not less likely to receive treatment when peers are assigned to treatment, holding own treatment assignment fixed. Third, we extend the idea of the IV exclusion restriction, which states that if an individual’s treatment assignment does not affect their own treatment status, it also does not effect their own outcome. Our irrelevance assumption extends this restriction to peers: If an individual’s treatment assignment does not affect own treatment status, then it also does not affect the peer’s treatment status. Similar identifying assumptions have been proposed for different parameters in different contexts, see for instance \citet{imai2021causal}.

The proposed set of assumptions apply to a wide range of IV settings in applied economics. The partial interference assumption only requires peers to be observable, which is reasonable in many economic settings where, for example, interactions happen within classrooms, but spillovers across them are negligible. In contrast, most recent studies on IV estimation with interference require specific experimental designs, such as two-stage randomization (\citet{kang2016peer,ditraglia2023identifying,imai2021causal}). \citet{hoshino2024causal} assumes access to an exposure mapping, defined as a low-dimensional summary statistic of the spillover effects, to allow for network interference of unknown form. \citet{ryu2024local} and \citet{acerenza2025bounds} propose identification results for more general experimental designs, but their results do not generalize to more than one peer. The same holds for \citet{vazquez2023causal}, who relies on restrictions on treatment effect heterogeneity when individuals have more than one peer. 

The proposed set of assumptions on the type of interference are also mild compared to the existing literature on IV estimation with interference. We allow for spillover effects of (1) the instrument on the treatment and (2) the treatment on the outcome. For instance, encouraging one individual to vaccinate may also encourage peers to get vaccinated, and one individual's vaccination may reduce peers' health risks. In contrast, \citet{kang2016peer} assume personalized encouragement which rules out the first spillover channel. \citet{ditraglia2023identifying} also restrict the second channel by assuming anonymous interactions, where the outcome only depends on peer treatment through the average treatment take-up across all peers. Both papers, together with \citet{vazquez2023causal}, also rely on one-sided noncompliance. Our framework only requires this assumption for the identification of spillover effects in the presence of multiple peers per individual.

Under our relatively mild identifying assumptions, the parameters of interest are partially identified. We construct bounds by relying on the irrelevance assumption introduced above together with generalizations of the monotone treatment response and monotone treatment selection assumptions of \cite{manski1997monotone} and \cite{manski2000monotone} to the IV setting with interference. These assumptions place no restrictions on the type or direction of interference. For instance, we allow spillover effects to be either positive (returns to scale) or negative (crowding out effects). In contrast, \citet{kormos2025interacting} bound an interaction effect in a more general setting, which has an interpretation of a spillover effect  in the context of interference, by assuming that and individual's outcome does not decrease when peers are treated. 

The outline of this paper is as follows. \Cref{sec:iv} discusses the IV framework with interference, defines the causal parameters of interest, and the identifying assumptions. \Cref{sec:pairs} discusses the identification of the parameters of interest in a setting in which each individual has one peer, and \Cref{sec:multiple} extend the results to a general number of peers. \Cref{sec:test} discusses the testable implications of the identifying assumptions. All proofs are deferred to the appendix.

\section{Instrumental variables with interference}\label{sec:iv}
Suppose we are interested in the causal effect of a binary treatment $D_i$ on an outcome $Y_i$, for individuals $i=1,\dots,n$. The treatment is potentially endogenous, and a binary instrument $Z_i$ is available to help to identify a causal effect. Individual $i$ has $m_i$ (potential) peers. The treatment status of the peers are collected in $D_{(i)}$, and their instruments in $Z_{(i)}$. Throughout this paper, we write $Z_{(i)}=z$ to indicate that all elements in $Z_{(i)}$ are equal to $z$, and similarly $D_{(i)}=d$ denotes that all elements in $D_{(i)}$ are equal to $d$. 

We use the potential outcome framework to define potential treatment status and outcome, while taking interference into account. Let $D_i(Z_i,Z_{(i)})$ denote the potential treatment status for individual $i$ and $D_{(i)}(Z_i,Z_{(i)})$ of $i$'s peers. The potential outcome for individual $i$ is denoted by $Y_i(D_i,D_{(i)},Z_i,Z_{(i)})$. If we account for interference, the instrumental variable assumptions are 
\begin{assumption}[Instrumental variable assumptions with interference]\label{ass:iv}\
\begin{enumerate}
    \item (Exclusion) $Y_i(d_i,d_{(i)},z_i,z_{(i)})=Y_i(d_i,d_{(i)})$. 
    \item (Independence) $Y_i(d_i,d_{(i)}),D_i(z_i,z_{(i)}) \perp (Z_i,Z_{(i)})$
    \item (Monotonicity) $D_i(1,z_{(i)}) \geq D_i(0,z_{(i)})$ and $D_{(i),j}(z_i,1) \geq D_{(i),j}(z_i,0)$.
\end{enumerate} 
\end{assumption}
Under the Stable Unit Treatment Value Assumption (SUTVA), $D_i(Z_i,Z_{(i)})=D_i(Z_i)$ and $Y_i(D_i,D_{(i)})=Y_i(D_i)$, and \Cref{ass:iv} boils down to the IV assumptions introduced by \citet{imbens1994identification}. They show that  
\begin{align}\label{eq:late}
    \frac{\mathbb{E}[Y_i|Z_i=1]-\mathbb{E}[Y_i|Z_i=0]}{\mathbb{E}[D_i|Z_i=1]-\mathbb{E}[D_i|Z_i=0]}=\mathbb{E}[Y_i(1)-Y_i(0)|D_i(1)>D_i(0)],
\end{align}
where the left-hand side is referred to as the IV estimand and the right-hand side the LATE for the individuals induced into treatment by the instrument, known as compliers. Under SUTVA, the instrument and treatment status of $i$'s peers do not affect $i$'s treatment and outcome respectively, and hence $Z_i$ only affects $Y_i$ through $D_i$.

\subsection{Causal parameters of interest}
In the presence of interference, the potential treatment status and outcome also depend on the instrument value and treatment status of the peers. The main parameter of interest is the direct effect, which is the causal effect of taking own treatment when peer's treatment status is fixed:
 \begin{align}\label{eq:tauD}
    \tau_{D}(d) = \mathbb{E}[Y_i(1,d)-Y_i(0,d)|D_i(1,d)> D_i(0,d),D_{(i)}(d,d)=d],
 \end{align}
for $d=0,1$, and where $\{D_i(1,d)\geq D_i(0,d)\}$ denotes the individuals induced into treatment by their own instrument.  When peer's treatment status is fixed at $d=0$ ($d=1$), $\{D_{(i)}(d,d)=d\}$ ensures that peers remain untreated (treated) when not assigned (assigned) to treatment. In the absence of interference, this effect does not depend on $d$, and equals the LATE in \eqref{eq:late}. Therefore, in case the LATE is of interest when SUTVA holds, the direct effects are of interest when there may be interference. 

With interference, there is also another parameter of interest. The indirect or spillover effect holds the own treatment status fixed, and changes the treatment status of the peers:
  \begin{align}\label{eq:tauS}
   \tau_{S}(d) =   \mathbb{E}[Y_i(d,D_{(i)}(d,1))-Y_i(d,0)|D_{(i)}(d,1) \neq D_{(i)}(d,0)=0,D_{i}(d,d)=d],
 \end{align}
 for $d=0,1$, and where $\{D_{(i)}(d,1) \neq D_{(i)}(d,0)=0\}$ denotes the peers from whom at least one is induced into treatment by their own instrument. With multiple peers, spillover effects can be present when only a subset of the peers is treated, and hence some elements in $D_{(i)}(d,1)$ may be zero. Under SUTVA, the potential outcomes only depend on $d$, and the spillover effect is therefore zero. With interference, spillover effects can be positive and improve the effectiveness of a treatment, or negative and reduce the overall effect.  

In treatment evaluation, both direct and spillover effects are important. For instance, $\tau_D(0)$ provides information on the treatment effect for the individuals induced into treatment, when their peers do not receive treatment. In case this effect is positive, negative spillover effects may still result in a negative evaluation of the whole treatment program. Estimating $\tau_S(0)$, which equals the spillover effects on the individuals who do not receive treatment, shows whether this may be the case. 

Next, the effects may inform policymakers on the scale of implementation of the treatment. Suppose individual $i$ has one peer. The difference between the effect of treating individual $i$ together with $i$'s peer and only treating individual $i$ equals $\tau_S(1)=\mathbb{E}[Y_i(1,1)-Y_i(0,0)|D_{(i)}(1,1) \neq D_{(i)}(1,0)=0,D_i(1,1)=1]-\mathbb{E}[Y_i(1,0)-Y_i(0,0)|D_{(i)}(1,1) \neq D_{(i)}(1,0)=0,D_i(1,1)=1]$. Hence, $\tau_S(1)$ provides insight into whether an individual benefits from full treatment adoption within the group of peers relative to treating only that individual. Similarly, $\tau_{D}(1) = \mathbb{E}[Y_i(1,1)-Y_i(0,0)|D_i(1,1)> D_i(0,1),D_{(i)}(1,1)=1]-\mathbb{E}[Y_i(0,1)-Y_i(0,0)|D_i(1,1)> D_i(0,1),D_{(i)}(1,1)=1]$, which shows whether an individual benefits from treatment if the peers are treated.

\subsection{Identifying assumptions with interference}
To identify the direct and spillover effects with interference, additional assumptions are required. First, we extend \Cref{ass:iv}.3 with a monotonicity assumption on the spillover effect of treatment assignment on treatment take-up:
\begin{assumption}[Monotonicity in the spillovers]\label{ass:monopeer}\ \\
     $D_i(z_i,1) \geq D_i(z_i,0)$ and $D_{(i),j}(1,z_{(i)}) \geq D_{(i),j}(0,z_{(i)})$.
\end{assumption}
The assumption states that an individual is not less likely to receive treatment when peers are assigned to treatment, holding own treatment assignment fixed. 

The identification problem can now be visualized by \Cref{fig:identification}. With interference, there are two possible spillover effects: the spillover effect of treatment assignment on treatment take-up ($Z_{i} \rightarrow D_{(i)}$) and the spillover effect of treatment take-up on the outcome ($D_{(i)} \rightarrow Y_i$). As a result, the treatment assignment of a noncomplier ($Z_i \nrightarrow D_i$) can still affect its outcome through the treatment status of its peers ($Z_i \rightarrow D_{(i)} \rightarrow Y_i$). 

\begin{figure}[t]
\centering
\caption{An overview of the possible effects on $Y_i$ with interference}
\label{fig:identification}
\begin{tikzpicture}[shorten >=2pt,node distance=7.5cm,on grid,auto]
   \node[] (K) {$Z_i$};   
   \node[right=50mm of K] (L) {$D_i$};
   \node[below=12.5mm of K] (N) {};
   \node[left=25mm of N] (J) {$Z$}; 
   \node[right=50mm of L] (M)  {$Y_i$};
   \node[below=25mm of K] (K1) {$Z_{(i)}$};
   \node[below=25mm of L] (L1) {$D_{(i)}$};
   \path[->]
    (J) edge [left]  node {} (K) 
    (K) edge [left]  node {} (L)
    (L) edge [left]  node {} (M)
    (K) edge [bend left, dashed]  node[below] {Exclusion}  (M)
    (J) edge [left]  node {} (K1)
    (K1) edge [left]  node {} (L1)
    (L1) edge [left, line width=.75mm]  node {} (M)
    (K) edge [left, dashed]  node {} (L1)
    (K1) edge [left, dashed] node[below=5mm] {Irrelevance}  node {} (L)
    (L) edge [bend left, line width=.75mm]  node {} (L1)
    (L1) edge [bend left, line width=.75mm]  node {} (L)
    (K1) edge [bend right, dashed]  node[pos=0.3,below=1mm] {Exclusion} (M)
    ;   
\end{tikzpicture}
\caption*{ \footnotesize
Notes: under \Cref{ass:iv}, $Z$ cannot be affected by any variable, $D_i$ and $D_{(i)}$ can only be affected by $Z_i$ and $Z_{(i)}$, and $Y$ can only be affected by $D_i$ and $D_{(i)}$. The dashed arrows represent the effects prevented by the stated assumptions.}
\end{figure}

To address this problem, we first extend the idea of the exclusion restriction in \Cref{ass:iv}.1 to the setting with interference: $Z_i$ only affects $Y_i$ through $D_i$, and therefore the outcome of noncompliers cannot be affected by $Z_i$.
\begin{assumption}[Irrelevance]\label{ass:irrelevance}\ \\
If $D_i(1,z_{(i)})=D_i(0,z_{(i)})$, then $D_{(i)}(1,z_{(i)})=D_{(i)}(0,z_{(i)})$.
\end{assumption}
This assumption states that if an individual's treatment is not affected by its own treatment assignment, then the treatment assignment of this individual has no effect on the peer's treatment status. \Cref{fig:identification} shows that under this irrelevance condition, $Z_i$ can only affect $D_{(i)}$ through $D_i$, and it follows that $Z_i$ can only affect $Y_i$ through $D_i$. Hence, the outcome of noncompliers is not affected by $Z_i$. \Cref{ass:irrelevance} still allows for both spillover channels, indicated by the bold arrows: the spillover effect of treatment assignment on treatment take-up ($Z_{i} \rightarrow D_i \rightarrow D_{(i)}$) and the spillover effect of treatment take-up on the outcome ($D_{(i)} \rightarrow Y_i$).

We propose partial identification results that are based on monotone treatment response (MTR) and monotone treatment selection (MTS) assumptions, which are commonly used to bound treatment effects in economics. Since each of our results depend on different MTR and MTS assumptions, and the plausibility of specific assumptions depend on the empirical setting, we state the exact assumptions in each result. Here we briefly discuss the general idea. We use MTR assumptions of the form
\begin{align}\label{eq:mtr}
    \text{ (MTR) } \mathbb{E}[Y(1,d)|I_i]\geq \mathbb{E}[Y(0,d)|I_i],
\end{align}
for $d=0,1$, and different compliance types $I_i$. Note that we invoke the assumption in expectation, instead of per individual, and that the assumption only applies to changes in own treatment: Taking up treatment does not decrease the outcome, holding treatment status of the peers fixed. Hence, we do not make any assumptions on how treatment take-up by peers affect an individuals outcome. These spillover effect could both be positive (returns to scale) or negative (crowding out effects). Our MTS assumption takes the following form:
\begin{align}\label{eq:mts}
   \text{ (MTS) } &\mathbb{E}[Y(d_i,d_{(i)})|D_i(0,0)>0]\geq
    \mathbb{E}[Y(d_i,d_{(i)})|D_i(1,0)+D_i(0,1)>0]\geq\\
    &\mathbb{E}[Y(d_i,d_{(i)})|D_i(1,1)>0]
    \geq\mathbb{E}[Y(d_i,d_{(i)})|D_i(1,1)=0],
\end{align}
for $d_i=0,1$, and $d_{(i)} \in \{0,1\}^{m_i}$. This implies that individuals who are more likely to receive treatment also have better potential outcomes.

\section{Interference within pairs}\label{sec:pairs}
In this section, we focus on interference within pairs. This setting only includes one peer $m_i=1$ for each individual $i$, which simplifies the exposition. The setting is also empirically relevant in itself, for instance for siblings, married couples or roommates. The next section extends the results to a general number of peers.  

The different possible potential treatment status $D_i(Z_i,Z_{(i)})$ define each individual's compliance type. \Cref{tab:types} lists all compliance types satisfying the monotonicity conditions in \Cref{ass:iv}.3 and \ref{ass:monopeer}.  
First, consider the three types also present in an IV setting with monotonicity but without interference. Always-takers ($A$) always receive treatment, never-takers ($N$) never receive treatment, and compliers ($C$) if and only if they are assigned to it.
With interference, three additional compliance types arise, which we refer to as spillover compliers. Social compliers ($S$) receive treatment as soon as themselves or their peer is assigned to it, peer compliers ($P$) receive treatment if and only if their peer is assigned to it, and group compliers ($G$) only receive the treatment when they and their peers are assigned to treatment. 

\begin{table}[t!] \setlength{\tabcolsep}{5pt}
  \centering \small
  \caption{Compliance types with one peer}
    \begin{tabular}{lccccrllccccc} 
    \toprule \toprule
    & \multicolumn{4}{c}{Compliance types}          &   &    & \multicolumn{6}{c}{Irrelevance exclusions with pairs} \\
             & \multicolumn{1}{l}{$D_i(1,1)$} & \multicolumn{1}{l}{$D_i(1,0)$} & \multicolumn{1}{l}{$D_i(0,1)$} & \multicolumn{1}{l}{$D_i(0,0)$} &       &       & $A_{(i)}$     & $S_{(i)}$     & $C_{(i)}$     & $P_{(i)}$     & $G_{(i)}$     & $N_{(i)}$ \\
    \cmidrule{1-5} \cmidrule{7-13}
    Always-taker         & 1     & 1     & 1     & 1     &       & $A_i$     &       & X     &       & X     & X     &   \\
    \textbf{Social complier}      & 1     & 1     & 1     & 0     &       & $S_i$     & X     &       &       & X     & X     & X \\
    Complier             & 1     & 1     & 0     & 0     &       & $C_i$     &       &       &       &       &       &   \\
    \textbf{Peer complier}        & 1     & 0     & 1     & 0     &       & $P_i$     & X     & X     &       & X     & X     & X \\
    \textbf{Group complier}       & 1     & 0     & 0     & 0     &       & $G_i$     & X     & X     &       & X     &       & X \\
    Never-taker          & 0     & 0     & 0     & 0     &       & $N_i$     &       & X     &       & X     & X     &   \\
    \bottomrule \bottomrule
    \end{tabular}%
  \label{tab:types}%
\end{table}%

\Cref{ass:irrelevance} does not restrict individual compliance types, but restricts certain combinations of compliance types within pairs. \Cref{tab:types} shows all exclusions within pairs. For instance, an always-taker is not affected by its own treatment assignment. It follows from the irrelevance condition that the treatment assignment of the always-taker has no effect on the peer’s treatment status ($D_{(i)}(1,1)=D_{(i)}(0,1)$ and $D_{(i)}(1,0)=D_{(i)}(0,0)$) and $\mathbb{P}[A_i,S_{(i)}]=\mathbb{P}[A_i,P_{(i)}]=\mathbb{P}[A_i,G_{(i)}]=0$.

\subsection{Direct effects}
Consider the average direct effect for individuals induced into treatment by their own instrument, when peer’s treatment status is fixed. From \Cref{tab:types} and \eqref{eq:tauD} follows that 
\begin{align}\label{eq:tauD0}
    \tau_D(0) = \mathbb{E}[Y_i(1,0)-Y_i(0,0)|S_iS_{(i)},S_iC_{(i)},C_iS_{(i)},C_iC_{(i)},C_iP_{(i)},C_iG_{(i)},C_iN_{(i)}],
\end{align}
which applies to compliers and social compliers; the individuals induced into treatment when the peers are not assigned to treatment. Similarly,
\begin{align}\label{eq:tauD1}
    \tau_D(1) = \mathbb{E}[Y_i(1,1)-Y_i(0,1)|C_iA_{(i)},C_iS_{(i)},C_iC_{(i)},C_iP_{(i)},C_iG_{(i)},G_iC_{(i)},G_iG_{(i)}],
\end{align}
applies to compliers and group compliers; the individuals induced into treatment when the peers are assigned to treatment. 

To formulate the identification results for the direct effects, we introduce some additional notation. Define $\Delta_{z_i'z_{(i)}'}^{z_iz_{(i)}}\mathbb{E}[A_i|Z]=\mathbb{E}[A_i|Z_i=z_i,Z_{(i)}=z_{(i)}]-\mathbb{E}[A_i|Z_i=z_i',Z_{(i)}=z_{(i)}']$. Define $ D_{(i)}^{\lor} = \prod_{j} (1-D_{(i),j})$ as indicator for none of the peers are treated, and $ D_{i(i)}^{\lor} = (1-D_i) D_{(i)}^{\lor}$ for no treatment for any of the peers and individual $i$. Denote $ D^{\land}_{(i)} = \prod_{j} D_{(i),j}$ as indicator for all peers are treated. Define $ D^{\land}_{i(i)} = D_iD^{\land}_{(i)}$ as indicator for all peers and $i$ treated.  Finally $Z_{i(i)}=(Z_i,Z_{(i)})$, with $Z_{i(i)}=z$ denoting that all elements in $Z_{i(i)}$ equal $z$.

\begin{lemma}[Partial identification direct effects with pairs]\label{lemma:direct}\ \\
Suppose Assumptions \ref{ass:iv}, \ref{ass:monopeer}, and \ref{ass:irrelevance} are satisfied. It holds that
\begin{enumerate}
    \item $L^{10}_{00}\leq \tau_D(0)\leq U^{10}_{00}$ if $\mathbb{P}[S_iS_{(i)},S_iC_{(i)},C_iS_{(i)},C_iC_{(i)},C_iP_{(i)},C_iG_{(i)},C_iN_{(i)}]>0$, 
\begin{align}
   L^{10}_{00} =& -\frac{\Delta^{10}_{00}\mathbb{E}[Y_i  D_{(i)}^{\lor}|Z]}{\Delta^{10}_{00}\mathbb{E}[D_{i(i)}^{\lor}|Z]}+\frac{\mathbb{E}[Y_iD_{i(i)}^{\lor}|Z_{i(i)}=1]}{\mathbb{E}[D_{i(i)}^{\lor}|Z_{i(i)}=1]}\frac{\Delta^{10}_{00}\mathbb{E}[ D_{(i)}^{\lor}|Z]}{\Delta^{10}_{00}\mathbb{E}[D_{i(i)}^{\lor}|Z]},
\end{align}
if $\mathbb{P}[N_iN_{(i)}]>0$ and $\mathbb{E}[Y_i(1,0)|S_i,C_i]\geq\mathbb{E}[Y_i(0,0)|S_i,C_i] \geq \mathbb{E}[Y_i(0,0)|N_i]$, and
\begin{align}
   U^{10}_{00} =&  -\frac{\Delta^{10}_{00}\mathbb{E}[Y_i  D_{(i)}^{\lor}|Z]}{\Delta^{10}_{00}\mathbb{E}[D_{i(i)}^{\lor}|Z]} +\frac{\Delta^{01}_{00}\mathbb{E}[Y_iD_iD_{(i)}^{\lor}|Z]}{\Delta^{01}_{00}\mathbb{E}[D_iD_{(i)}^{\lor}|Z]}\frac{\Delta^{10}_{00}\mathbb{E}[ D_{(i)}^{\lor}|Z]}{\Delta^{10}_{00}\mathbb{E}[D_{i(i)}^{\lor}|Z]},
\end{align}
if $\mathbb{P}[A_iC_{(i)}]>0$ and $\mathbb{E}[Y_i(1,0)|A_i] \geq \mathbb{E}[Y_i(1,0)|C_i,S_i]$.
\item $L^{11}_{01} \leq \tau_D(1) \leq U^{11}_{01}$ if $\mathbb{P}[C_iA_{(i)},C_iS_{(i)},C_iC_{(i)},C_iP_{(i)},C_iG_{(i)},G_iC_{(i)},G_iG_{(i)}]>0$, 
\begin{align}
   L^{11}_{01} =& \frac{\Delta^{11}_{01}\mathbb{E}[Y_i  D_{(i)}^{\land}|Z]}{\Delta^{11}_{01}\mathbb{E}[D_{i(i)}^{\land}|Z]}- \frac{\mathbb{E}[Y_iD_{i(i)}^{\land}|Z_{i(i)}=0]}{\mathbb{E}[D_{i(i)}^{\land}|Z_{i(i)}=0]}\frac{\Delta^{11}_{01}\mathbb{E}[ D_{(i)}^{\land}|Z]}{\Delta^{11}_{01}\mathbb{E}[D_{i(i)}^{\land}|Z]},
\end{align}
if $\mathbb{P}[A_iA_{(i)}]>0$ and $\mathbb{E}[Y_i(1,1)|A_i]\geq\mathbb{E}[Y_i(0,1)|A_i] \geq \mathbb{E}[Y_i(0,1)|C_i,G_i]$, and
\begin{align}
   U^{11}_{01} =&  \frac{\Delta^{11}_{01}\mathbb{E}[Y_i  D_{(i)}^{\land}|Z]}{\Delta^{11}_{01}\mathbb{E}[D_{i(i)}^{\land}|Z]}-\frac{\Delta^{11}_{10}\mathbb{E}[Y_i(1-D_i)D_{(i)}^{\land}|Z]}{\Delta^{11}_{10}\mathbb{E}[(1-D_i)D_{(i)}^{\land}|Z]}\frac{\Delta^{11}_{01}\mathbb{E}[ D_{(i)}^{\land}|Z]}{\Delta^{11}_{01}\mathbb{E}[D_{i(i)}^{\land}|Z]},
\end{align}
if $\mathbb{P}[N_iC_{(i)}]>0$ and $\mathbb{E}[Y_i(0,1)|C_i,G_i] \geq \mathbb{E}[Y_i(0,1)|N_i]$.
\end{enumerate}
\end{lemma}

The difference between the bounds is a function of the proportion of spillover compliers.
More specifically, $U^{10}_{00}-L^{10}_{00}$ decreases in $\mathbb{P}[S_iS_{(i)},C_iS_{(i)},C_iP_{(i)}]/\mathbb{P}[C_iC_{(i)},C_iG_{(i)},C_iN_{(i)},S_iC_{(i)}]$ and $U^{11}_{01}-L^{11}_{01}$ in $\mathbb{P}[G_iG_{(i)},C_iG_{(i)},C_iP_{(i)}]/\mathbb{P}[C_iA_{(i)},C_iS_{(i)},C_iC_{(i)},G_iC_{(i)}]$. Therefore, $\tau_D(0)$ ($\tau_D(1)$) is point identified if interference is solely due to group (social) compliers. 
In the absence of any interference, both parameters are point identified.

For the direct effects in \eqref{eq:tauD0} and \eqref{eq:tauD1} to exist, the sets of corresponding compliance types has to be nonempty. These conditions are both satisfied when $\mathbb{P}[C_iS_{(i)},C_iC_{(i)},C_iP_{(i)},C_iG_{(i)}]>0$, which is similar to the relevance assumption in standard IV. 
The bounds in \Cref{lemma:direct} rely on potential outcomes of always-takers and never-takers paired with the same compliance type ($A_iA_{(i)}$ and $N_iN_{(i)}$) or a complier ($A_iC_{(i)}$ and $N_iC_{(i)}$). These pairs exist under two-sided noncompliance: some individuals never take treatment even when assigned, while some individuals always take treatment even when not assigned. We discuss the extension of \Cref{lemma:direct} to one-sided noncompliance in \Cref{sec:multiple_spillover}.
The potential outcomes of these always-takers and never-takers are used to bound the potential outcomes of individual's paired with social, peer, or group compliers, according to the MTR and MTS assumptions defined in \eqref{eq:mtr} and \eqref{eq:mts}.  

\subsection{Spillover effects}
The average spillover effects in \eqref{eq:tauS} can be written as
\begin{align}\label{eq:tauS0}
    \tau_S(0) = \mathbb{E}[Y_i(0,1)-Y_i(0,0)|S_iS_{(i)},C_iS_{(i)},S_iC_{(i)},C_iC_{(i)},P_iC_{(i)}, G_iC_{(i)},N_iC_{(i)}],
\end{align}
which applies to individuals with a complier or social complier peer, while not receiving treatment themselves. Similarly,
\begin{align}\label{eq:tauS1}
    \tau_S(1) = \mathbb{E}[Y_i(1,1)-Y_i(1,0)| A_iC_{(i)},S_iC_{(i)},C_iC_{(i)},P_iC_{(i)},G_iC_{(i)},C_iG_{(i)},G_iG_{(i)}],
\end{align}
applies to individuals with a complier or group complier peer, while receiving treatment themselves. Both spillover effects are partially identified:

\begin{lemma}[Partial identification spillover effects with pairs]\label{lemma:spillover}\ \\
Suppose Assumptions \ref{ass:iv}, \ref{ass:monopeer}, and \ref{ass:irrelevance} are satisfied. It holds that
\begin{enumerate}
    \item $L^{01}_{00} \leq \tau_S(0) \leq U^{01}_{00}$ if $\mathbb{P}[S_iS_{(i)},C_iS_{(i)},S_iC_{(i)},C_iC_{(i)},P_iC_{(i)}, G_iC_{(i)},N_iC_{(i)}]>0$,
\begin{align}
    L^{01}_{00} =& -\frac{\Delta^{01}_{00}\mathbb{E}[Y_i(1-D_i)|Z]}{\Delta^{01}_{00}\mathbb{E}[D_{i(i)}^{\lor}|Z]}-\frac{\Delta^{11}_{10}\mathbb{E}[Y_i(1-D_i)D_{(i)}|Z]}{\Delta^{11}_{10}\mathbb{E}[(1-D_i)D_{(i)}|Z]}\frac{\Delta^{01}_{00}\mathbb{E}[D_{i}|Z]}{\Delta^{01}_{00}\mathbb{E}[D_{i(i)}^{\lor}|Z]},
\end{align}
if $\mathbb{P}[N_iC_{(i)}]>0$ and $\mathbb{E}[Y_i(0,1)|S_i,P_i] \geq \mathbb{E}[Y_i(0,1)|N_i]$, and
\begin{align}
    U^{01}_{00} =& -\frac{\Delta^{01}_{00}\mathbb{E}[Y_i(1-D_i)|Z]}{\Delta^{01}_{00}\mathbb{E}[D_{i(i)}^{\lor}|Z]}-\frac{\mathbb{E}[Y_iD_{i(i)}^{\land}|Z_{i(i)}=0]}{\mathbb{E}[D_{i(i)}^{\land}|Z_{i(i)}=0]}\frac{\Delta^{01}_{00}\mathbb{E}[D_{i}|Z]}{\Delta^{01}_{00}\mathbb{E}[D_{i(i)}^{\lor}|Z]},
\end{align}
if $\mathbb{P}[A_iA_{(i)}]>0$ and $\mathbb{E}[Y_i(1,1)|A_i] \geq \mathbb{E}[Y_i(0,1)|A_i] \geq \mathbb{E}[Y_i(0,1)|S_i,P_i]$.
\item $L^{11}_{10}\leq \tau_S(1) \leq U^{11}_{10}$, if $\mathbb{P}[ A_iC_{(i)},S_iC_{(i)},C_iC_{(i)},P_iC_{(i)},G_iC_{(i)},C_iG_{(i)},G_iG_{(i)}]>0$,
\begin{align}
    L^{11}_{10}=& \frac{\Delta^{11}_{10} \mathbb{E}[Y_iD_i|Z] }{\Delta^{11}_{10} \mathbb{E}[D_{i(i)}^\land|Z]}-\frac{\Delta^{01}_{00}\mathbb{E}[Y_iD_iD_{(i)}^{\lor}|Z]}{\Delta^{01}_{00}\mathbb{E}[D_iD_{(i)}^{\lor}|Z]}\frac{\Delta^{11}_{10} \mathbb{E}[D_{i}|Z]}{\Delta^{11}_{10} \mathbb{E}[D_{i(i)}^\land|Z]},
\end{align}
if $\mathbb{P}[A_iC_{(i)}]$ and $\mathbb{E}[Y_i(1,0)|A_i] \geq \mathbb{E}[Y_i(1,0)|P_i,G_i]$, and
\begin{align}
    U^{11}_{10}=& \frac{\Delta^{11}_{10} \mathbb{E}[Y_iD_i|Z] }{\Delta^{11}_{10} \mathbb{E}[D_{i(i)}^\land|Z]}-\frac{\mathbb{E}[Y_iD_{i(i)}^{\lor}|Z_{i(i)}=1]}{\mathbb{E}[D_{i(i)}^{\lor}|Z_{i(i)}=1]}\frac{\Delta^{11}_{10} \mathbb{E}[D_{i}|Z]}{\Delta^{11}_{10} \mathbb{E}[D_{i(i)}^\land|Z]},
\end{align}
if $\mathbb{P}[N_iN_{(i)}]$ and $\mathbb{E}[Y_i(1,0)|P_i,G_i] \geq \mathbb{E}[Y_i(0,0)|P_i,G_i] \geq \mathbb{E}[Y_i(0,0)|N_i]$.
\end{enumerate}
\end{lemma}

Similar as to \Cref{lemma:direct}, the difference between the bounds depend on the proportion of spillover compliers: $U^{01}_{00}-L^{01}_{00}$ decreases in $\mathbb{P}[S_iS_{(i)},S_iC_{(i)},P_iC_{(i)}]/\mathbb{P}[C_iS_{(i)},C_iC_{(i)}, G_iC_{(i)},N_iC_{(i)}]$ and $U^{11}_{10}-L^{11}_{10}$ in $\mathbb{P}[P_iC_{(i)},G_iC_{(i)},G_iG_{(i)}]/\mathbb{P}[A_iC_{(i)},S_iC_{(i)},C_iC_{(i)},C_iG_{(i)}]$. Therefore, $\tau_S(0)$ ($\tau_S(1)$) is point identified if interference is solely due to group (social) compliers. In the absence of any interference, both parameters are point identified.

\Cref{lemma:spillover} also relies on similar assumptions as \Cref{lemma:direct}. 
The relevance assumption required here switches the compliance types between individual $i$ and $i$'s peer: $\mathbb{P}[S_iC_{(i)},C_iC_{(i)},P_iC_{(i)}, G_iC_{(i)}]>0$. The MTR and MTS assumptions apply to $P_i$ instead of $C_i$, since the peer is induced into treatment here, but are identical otherwise. Both results require the same pairs with always-takers and never-takers to exist. 

\section{Interference with multiple peers}\label{sec:multiple}
The results in \Cref{sec:pairs} can be extended to $m_i>1$ peers. However, if we let the number of peers an endogenous choice, identification becomes infeasible for two reasons. First, although the treatment assignment $Z=\{Z_i\}_{i=1}^n$ to all individuals is random, individuals with more peers are expected to have more treated peers. The treatment assignment of the peers $Z_{(i)}$ is not only a function of $Z$, but also of the individuals selected to be individual $i$'s peers.  
This violates \Cref{ass:iv}.2 if individuals with more peers systematically differ in their outcomes from individuals with less peers. \citet{borusyak2023nonrandom} discuss this identification problem in more detail. 
Second, for the identification problem to be solvable, the number of peers need to be small relative to the sample size. Without restricting the number of peers, each individual potentially has $n-1$ peers and therefore $2^n$ potential treatment and outcome values.

\begin{assumption}[Partial interference]\label{ass:partial}\ \\
    Suppose we observe $n$ individuals, and each individual $i$ has $m_i$ peers. It holds that $m_i=m$ with $m<<n$. 
\end{assumption}

\Cref{ass:partial} addresses both identification issues with multiple peers; the number of peers is equal and fixed for each individual. This reduces the number of potential treatment and outcome values to $2^m$. For our parameters of interest in \eqref{eq:tauD} and \eqref{eq:tauS}, we either assign all peers to treatment or none. It follows that we only need to observe these two types of treatment assignment for the peers, instead of $2^m$ different assignment combinations. 
In practice, we use \Cref{ass:partial} to ensure that we only compare individuals with the same group size for estimating the treatment effects in \eqref{eq:tauD} and \eqref{eq:tauS}. However, treatment effect estimates may be averaged across varying group sizes, as for example in \citet{imai2021causal}. The averaged effects can be interpreted as the direct effect when all or none of the peers are treated, or the spillover effect from all peers being treated, while the number of peers varies. 

Allowing for multiple peers complicates the use of the compliance type notation for the peers. For instance, the potential treatment status notation $D_{(i)}(0,1)\neq D_{(i)}(0,0)=0$ in \eqref{eq:tauS} translates to either $S_{(i)}$ or $C_{(i)}$ with one peer. With multiple pairs, this includes any group of peers containing at least one (social) complier and no always-takers, resulting in too many possible combinations to enumerate. Hence, we only use the compliance type notation for individual $i$ in this section, and use the potential treatment status notation for the peers of $i$. As an exception, we use $N_{(i)}$  ($A_{(i)}$) when all peers are never- (always-)takers. 

\subsection{Direct effects}\label{sec:multiple_direct}
The bounds in \Cref{lemma:direct} directly generalize to multiple peers:

\begin{theorem}[Partial identification direct effects]\label{theorem:direct}\ \\
Suppose Assumptions \ref{ass:iv}, \ref{ass:monopeer}, \ref{ass:irrelevance},  and \ref{ass:partial} are satisfied. It holds that
\begin{enumerate}[leftmargin=0.5cm]
    \item $L^{10}_{00}\leq \tau_D(0)\leq U^{10}_{00}$ with $L^{10}_{00}$ and $U^{10}_{00}$ defined in \Cref{lemma:direct}, and 
    \begin{enumerate}[leftmargin=0cm]
        \item[] $\mathbb{P}[\{S_i,C_i\}\times \{D_{(i)}(0,0)=0\}]>0$ for $\tau_D(0)$ to exist;
        \item[] $\mathbb{P}[N_iN_{(i)}]>0$ and $\mathbb{E}[Y_i(1,0)|S_i,C_i]\geq\mathbb{E}[Y_i(0,0)|S_i,C_i] \geq \mathbb{E}[Y_i(0,0)|N_i]$ for $L^{10}_{00}$;
        \item[] $\mathbb{P}[A_i\times\{D_{(i)}(0,1)\neq D_{(i)}(0,0)=0\}]>0$ and $\mathbb{E}[Y_i(1,0)|A_i] \geq \mathbb{E}[Y_i(1,0)|S_i,C_i]$ for $U^{10}_{00}$.
    \end{enumerate}
    \item $L^{11}_{01} \leq \tau_D(1) \leq U^{11}_{01}$ with $L^{11}_{01}$ and $U^{11}_{01}$ defined in \Cref{lemma:direct}, and 
    \begin{enumerate}[leftmargin=0cm]
        \item[] $\mathbb{P}[\{C_i,G_i\}\times \{D_{(i)}(1,1)=1\}]>0$ for $\tau_D(1)$ to exist;
        \item[]  $\mathbb{P}[A_iA_{(i)}]>0$ and $\mathbb{E}[Y_i(1,1)|A_i]\geq\mathbb{E}[Y_i(0,1)|A_i] \geq \mathbb{E}[Y_i(0,1)|C_i,G_i]$ for $L^{11}_{01}$.
        \item[] $\mathbb{P}[N_i\times \{D_{(i)}(1,0)\neq D_{(i)}(1,1)=1\}]>0$ and $\mathbb{E}[Y_i(0,1)|C_i,G_i] \geq \mathbb{E}[Y_i(0,1)|N_i]$ for $U^{11}_{01}$.
    \end{enumerate}
\end{enumerate}
\end{theorem}

Similar to \Cref{lemma:direct}, $\tau_D(0)$ and $\tau_D(1)$ are point-identified in the absence of social and peer compliers, or the absence of group and peer compliers, respectively. The difference between the bounds decreases in the proportion of these compliers. The direct effects exist if there are compliers who do not have any always-takers and never-takers as peers. The bounds require always- (never-) takers with only always- (never-) takers as peers, and always-(never-) takers with at least one social complier or complier but no always-takers (group complier or complier but no never-takers) as peer. The potential outcomes of these compliance types are used in the MTR and MTS assumptions, which impose that taking own treatment does not decrease outcome, holding all peers fixed to treatment or no treatment, and that always-takers are more likely to have better outcomes than compliers, which in turn are more likely to have better outcomes than never-takers.

\subsection{Spillover effects}\label{sec:multiple_spillover}
The identification results for the spillover effects in \Cref{lemma:spillover} do not directly generalize to the multiple peer setting. 
With one peer, we use the peer's instrument to partially identify a spillover effect, while the individual's own instrument is fixed; otherwise we cannot distinguish the effect of the individual's own instrument from the peer's instruments on the individual's treatment. With multiple peers, any peer's instrument, or any combination of peer's instruments, could have induced a spillover effect. This requires to take all possible combinations of peer instruments into account, which becomes infeasible when the number of peers increases. 

We show identification results for the spillover effects with multiple peers under a one-sided noncompliance assumption:
\begin{assumption}[One-sided noncompliance]\label{ass:osnc}\ \\
$\mathbb{P}[D_i=1|Z_i=0]=0$.
\end{assumption}
Because of the challenging identification problem, this assumption is common for the identification of spillover effects with multiple peers \citep{kang2016peer,ditraglia2023identifying,kormos2025interacting,vazquez2023causal}. One-sided noncompliance is also common in treatment evaluation, with many empirical settings in which individuals who are not assigned to treatment are unable to obtain treatment.

Since \Cref{ass:osnc} sets $\mathbb{P}[D_i(0,1)=1]=\mathbb{P}[D_i(0,0)=1]=0$, it follows from \Cref{tab:types} that always-takers, social compliers, and peer compliers are excluded. The spillover effects in \eqref{eq:tauS} therefore simplify to
\begin{align}
    \tau_S(0)=\mathbb{E}[Y_i(0,D_{(i)}(0,1))-Y_i(0,0)|D_{(i)}(0,1)\neq 0],
\end{align}
where \Cref{ass:osnc} excludes the always-takers and hence $D_{(i)}(0,0)=D_i(0,0)=0$. Similarly,
\begin{align}
    \tau_S(1)=\mathbb{E}[Y_i(1,D_{(i)}(1,1))-Y_i(1,0)|D_{(i)}(1,1)\neq 0,D_i(1,1)=1],
\end{align}
where \Cref{ass:osnc} excludes always-takers, social compliers, and peer compliers and $D_{(i)}(1,0)=0$.

\begin{theorem}[Identification spillover effects]\label{theorem:spillover}\ \\
Suppose Assumptions \ref{ass:iv}, \ref{ass:monopeer}, \ref{ass:irrelevance},  \ref{ass:partial} and \ref{ass:osnc} are satisfied. It holds that
\begin{enumerate}
\item $\tau_S(0) = -\Delta^{01}_{00}\mathbb{E}[Y_i(1-D_i)|Z]/\Delta^{01}_{00}\mathbb{E}[D_{i(i)}^{\lor}|Z]$ if $\mathbb{P}[D_{(i)}(0,1)\neq 0]>0$.
\item $ L^{11}_{10} \leq \tau_S(1) \leq U^{11}_{10}$ if $\mathbb{P}[\{C_i,G_i\}\times\{D_{(i)}(1,1)\neq 0\}]>0$, 
\begin{align}
    L^{11}_{10} =& \frac{\Delta^{11}_{10}\mathbb{E}[Y_iD_i|Z]}{\Delta^{11}_{10}\mathbb{E}[D_i(1- D_{(i)}^{\lor})|Z]}-\frac{\Delta^{11}_{10}\mathbb{E}[Y_iD_i D_{(i)}^{\lor}|Z]}{\Delta^{11}_{10}\mathbb{E}[D_i D_{(i)}^{\lor}|Z]}\frac{\Delta^{11}_{10}\mathbb{E}[D_i|Z]}{\Delta^{11}_{10}\mathbb{E}[D_i(1- D_{(i)}^{\lor})|Z]},
\end{align}
if $\mathbb{P}[C_i \times \{D_{(i)}(1,1)\neq0\}]>0$ and $\mathbb{E}[Y_i(1,0)|C_i] \geq \mathbb{E}[Y_i(1,0)|G_i]$, and
\begin{align}
    U^{11}_{10} =& \frac{\Delta^{11}_{10}\mathbb{E}[Y_iD_i|Z]}{\Delta^{11}_{10}\mathbb{E}[D_i(1- D_{(i)}^{\lor})|Z]}-\frac{\mathbb{E}[Y_iD_{i(i)}^{\lor}|Z_{i(i)}=1]}{\mathbb{E}[D_{i(i)}^{\lor}|Z_{i(i)}=1]}\frac{\Delta^{11}_{10}\mathbb{E}[D_i|Z]}{\Delta^{11}_{10}\mathbb{E}[D_i(1- D_{(i)}^{\lor})|Z]},
\end{align}
if $\mathbb{P}[N_iN_{(i)}]>0$ and $\mathbb{E}[Y_i(1,0)|G_i] \geq \mathbb{E}[Y_i(0,0)|G_i] \geq \mathbb{E}[Y_i(0,0)|N_i]$.
\end{enumerate}
\end{theorem}

Under \Cref{ass:osnc}, the only spillover complier is the group complier. Since an instrument switch from $Z_i=0$ and $Z_{(i)}=0$ to $Z_i=0$ and $Z_{(i)}=1$ does not induce the group complier, $\tau_S(0)$ is now point identified. If the group complier is absent, both parameters are point identified. 

\section{Testable implications}\label{sec:test}
When SUTVA holds,  the IV method introduced by \citet{imbens1994identification} point-identifies the LATE, which equals both direct effects, and the spillover effects equal zero. When there is interference, this paper shows partial identification results for the direct and spillover effects. Hence, in the absence of interference, standard IV approaches are to be preferred.
The next proposition presents necessary conditions that can be used for falsification tests of the SUTVA assumption.
\begin{proposition}[Necessary conditions SUTVA]\label{prop:sutva}\ \\
    Suppose Assumptions \ref{ass:iv}, \ref{ass:monopeer}, and \ref{ass:partial} are satisfied. Under SUTVA it holds that
    \begin{align}
        \Delta^{01}_{00}\mathbb{E}[D_i|Z]=\mathbb{P}[S_i,P_i]=0 \text{ and } \Delta^{11}_{10}\mathbb{E}[D_i|Z]=\mathbb{P}[P_i,G_i]=0.
    \end{align}  
\end{proposition}

In the absence of peer compliers, \Cref{prop:sutva} can also be used to find the type of spillover compliers. \citet{vazquez2023causal} assumes that $\mathbb{P}[P_i]=0$ by imposing an additional monotonicity assumption $D_i(1,0)\geq D_i(0,1)$. The peer compliers are also absent with one-sided noncompliance, which can be verified by the condition $\mathbb{E}[D_i|Z_i=0,Z_{(i)}=1]=\mathbb{P}[A_i,S_i,P_i]=0$. 

We also provide necessary conditions for the irrelevance condition in \Cref{ass:irrelevance}:
\begin{proposition}[Necessary conditions irrelevance]\label{prop:irrelevance}\ \\
    Suppose Assumptions \ref{ass:iv}, \ref{ass:monopeer}, and \ref{ass:partial} are satisfied. Under \Cref{ass:irrelevance} it holds that for $d=0,1$:
    \begin{align}
        \Delta^{1d}_{0d} \mathbb{E}[D_i D_{(i)}^\lor|Z]\geq 0 &\text{ and } \Delta^{1d}_{0d} \mathbb{E}[(1-D_i) D_{(i)}^\land|Z]\leq 0,\\
        \Delta^{d1}_{d0} \mathbb{E}[D_i D_{(i)}^\lor|Z]\leq 0 &\text{ and } \Delta^{d1}_{d0} \mathbb{E}[(1-D_i) D_{(i)}^\land|Z]\geq 0.
    \end{align}  
\end{proposition}
These conditions are similar to the ones derived by \citet{hoshino2024causal} in the setting of an exposure mapping.

The proportion of the compliance types required for each identification result can also be identified from the data. In fact, they are identified by the denominators in the bounds, which have to be nonzero for the bounds to exist. For instance, the denominators in $L^{10}_{00}$ and $U^{10}_{00}$ in \Cref{lemma:direct} equal $\Delta^{10}_{00}\mathbb{E}[D_{i(i)}^{\lor}|Z]=\mathbb{P}[S_iS_{(i)},S_iC_{(i)},C_iS_{(i)},C_iC_{(i)},C_iP_{(i)},C_iG_{(i)},C_iN_{(i)}]$, $\mathbb{E}[D_{i(i)}^{\lor}|Z_{i(i)}=1]=\mathbb{P}[N_iN_{(i)}]$ and $\Delta^{01}_{00}\mathbb{E}[D_iD_{(i)}^{\lor}|Z]=\mathbb{P}[A_iC_{(i)}]$. 
In case the compliance types required for the MTR and MTS assumptions do not exist, the corresponding potential outcomes can be replaced by the maximum and minimum possible values for $Y$, or use alternative compliance types, as discussed in the appendix. 

\section{Conclusion}
This paper defines causal parameters of interest with a clear policy relevant interpretation in the context of instrumental variable estimation with interference. We provide partial identification results for these parameters, both in settings in which individuals have one peer and multiple peers. Testable necessary conditions for the identifying assumptions are proposed. We are working on an outline of the implementation details of the proposed methods, and empirical illustrations.  

\bibliographystyle{chicago}
\bibliography{Library.bib}

\clearpage
\appendix

\newpage
\setcounter{page}{1}
\setcounter{equation}{0}
\setcounter{table}{0}
\setcounter{figure}{0}

\section{Proof \Cref{lemma:direct} and \Cref{theorem:direct}}

\subsection{First stages in bounds $\tau_D(0)$}
Derive expressions for $\Delta^{10}_{00}\mathbb{E}[ D_{(i)}^{\lor}|Z]$, $\Delta^{10}_{00}\mathbb{E}[ D_{i(i)}^{\lor}|Z]$, $\Delta^{01}_{00}\mathbb{E}[D_iD_{(i)}^{\lor}|Z]$, and $\mathbb{E}[D_{i(i)}^{\lor}|Z_{i(i)}=1]$.
\begin{align}\label{eq:fstauD0}
    \Delta^{10}_{00}\mathbb{E}[ D_{(i)}^{\lor}|Z] =& \mathbb{E}[D_{(i)}^{\lor}(1,0)-D_{(i)}^{\lor}(0,0)]\\
    =&-\mathbb{P}[D_{i}(1,0) > D_{i}(0,0),D_{(i)}(1,0) \neq D_{(i)}(0,0)=0],
\end{align}
where we use \Cref{ass:iv}.2, and subsequently that $D_{(i),j}(1,0) \geq D_{(i),j}(0,0)$ for all peers $j$ according to \Cref{ass:monopeer}, and if $D_{(i)}(1,0) \neq D_{(i)}(0,0)$, it follows from \Cref{ass:irrelevance} and \Cref{ass:iv}.3 that $D_{i}(1,0) > D_{i}(0,0)$.
\begin{align}
    \Delta^{10}_{00}\mathbb{E}[ D_{i(i)}^{\lor}|Z] =& \mathbb{E}[D_{i(i)}^{\lor}(1,0)-D_{i(i)}^{\lor}(0,0)]\\ 
    =& -\mathbb{P}[D_{i}(1,0) > D_{i}(0,0),D_{(i)}(1,0)=D_{(i)}(0,0)=0]- \notag\\
     &\mathbb{P}[D_{i}(1,0) > D_{i}(0,0),D_{(i)}(1,0)\neq D_{(i)}(0,0)=0],
\end{align}
where we use \Cref{ass:iv}.2, and subsequently that $D_{(i),j}(1,0) \geq D_{(i),j}(0,0)$ for all peers $j$ according to \Cref{ass:monopeer}. If $D_{(i)}(1,0) = D_{(i)}(0,0)$, we require $D_{i}(1,0) \neq D_{i}(0,0)$ for a nonzero expression. If $D_{(i)}(1,0) \neq D_{(i)}(0,0)$, it follows from \Cref{ass:irrelevance} that $D_{i}(1,0) \neq D_{i}(0,0)$.  In both cases,  $D_{i}(1,0) > D_{i}(0,0)$ due to  \Cref{ass:iv}.3.
\begin{align}\label{eq:AC}
    \Delta^{01}_{00}\mathbb{E}[D_iD_{(i)}^{\lor}|Z] =& \mathbb{E}[D_i(0,1)D_{(i)}^{\lor}(0,1)-D_i(0,0)D_{(i)}^{\lor}(0,0)]\\
    =& -\mathbb{P}[D_i(0,1)=D_i(0,0)=1,D_{(i)}(0,1) \neq D_{(i)}(0,0)=0],
\end{align}
where we use \Cref{ass:iv}.2, and subsequently that $D_{(i),j}(0,1) \geq D_{(i),j}(0,0)$ for all peers $j$ according to \Cref{ass:iv}.3. It follows from \Cref{ass:irrelevance} that we require $D_{(i)}(0,1) \neq D_{(i)}(0,0)$ and $D_i(0,0)=1$ for a nonzero expression. Finally we use \Cref{ass:monopeer}. Using \Cref{ass:iv}.2, we have
\begin{align}\label{eq:NN}
    \mathbb{E}[D_{i(i)}^{\lor}|Z_{i(i)}=1] = \mathbb{E}[D_{i(i)}^{\lor}(1,1)]=\mathbb{P}[D_i(1,1)=0,D_{(i)}(1,1)=0]. 
\end{align}

\subsection{Reduced forms in bounds $\tau_D(0)$}
Derive expressions for $\Delta^{10}_{00}\mathbb{E}[Y_i  D_{(i)}^{\lor}|Z]$, $\Delta^{01}_{00}\mathbb{E}[Y_iD_iD_{(i)}^{\lor}|Z]$, and $\mathbb{E}[Y_iD_{i(i)}^{\lor}|Z_{i(i)}=1]$.
\begin{align}
    \Delta^{10}_{00}\mathbb{E}[Y_i  D_{(i)}^{\lor}|Z] =& \mathbb{E}[Y_i(D_i(1,0),0)D_{(i)}^{\lor}(1,0)-Y_i(D_i(0,0),0)D_{(i)}^{\lor}(0,0)] \\
    =& \mathbb{E}[Y_i(1,0)-Y_i(0,0)|D_{i}(1,0) > D_{i}(0,0),D_{(i)}(1,0)=D_{(i)}(0,0)=0] \notag\\
     & \times\mathbb{P}[D_{i}(1,0) > D_{i}(0,0),D_{(i)}(1,0)=D_{(i)}(0,0)=0]-\notag\\
     & \mathbb{E}[Y_i(0,0)|D_{i}(1,0) > D_{i}(0,0),D_{(i)}(1,0)\neq D_{(i)}(0,0)=0]\notag\\
     & \times\mathbb{P}[D_{i}(1,0) > D_{i}(0,0),D_{(i)}(1,0)\neq D_{(i)}(0,0)=0],
\end{align}
where we first use \Cref{ass:iv}.1 and \ref{ass:iv}.2, and use that $D_{(i)}^{\lor}=0$ if there is a peer $j$ with $D_{(i),j}=1$. Second, we use the same arguments as for the derivation of $\Delta^{10}_{00}\mathbb{E}[ D_{i(i)}^{\lor}|Z]$.  Note that we can rewrite $\Delta^{10}_{00}\mathbb{E}[Y_i  D_{(i)}^{\lor}|Z]$ to
\begin{align}
    \Delta^{10}_{00}\mathbb{E}[Y_i  D_{(i)}^{\lor}|Z] =& \mathbb{E}[Y_i(1,0)-Y_i(0,0)|D_{i}(1,0) > D_{i}(0,0),D_{(i)}(0,0)=0]\notag\\
     & \times\mathbb{P}[D_{i}(1,0) > D_{i}(0,0),D_{(i)}(0,0)=0]-\notag\\
     & \mathbb{E}[Y_i(1,0)|D_{i}(1,0) > D_{i}(0,0),D_{(i)}(1,0)\neq D_{(i)}(0,0)=0]\notag\\
     & \times\mathbb{P}[D_{i}(1,0) > D_{i}(0,0),D_{(i)}(1,0)\neq D_{(i)}(0,0)=0].
\end{align}
\begin{align}\label{eq:YAC}
    \Delta^{01}_{00}\mathbb{E}[Y_iD_iD_{(i)}^{\lor}|Z]=&\mathbb{E}[Y_i(1,0)\left(D_i(0,1)D_{(i)}^{\lor}(0,1)-D_i(0,0)D_{(i)}^{\lor}(0,0)\right)]\\
    =& -\mathbb{E}[Y_i(1,0)|D_i(0,1)=D_i(0,0)=1,D_{(i)}(0,1) \neq D_{(i)}(0,0)=0]
    \notag\\&\times\mathbb{P}[D_i(0,1)=D_i(0,0)=1,D_{(i)}(0,1) \neq D_{(i)}(0,0)=0],
\end{align}
where we first use \Cref{ass:iv}.1 and \ref{ass:iv}.2, and use that $D_{(i)}^{\lor}=0$ if there is a peer $j$ with $D_{(i),j}=1$. Second, we use the same arguments as for the derivation of $\Delta^{01}_{00}\mathbb{E}[D_iD_{(i)}^{\lor}|Z]$.
\begin{align}\label{eq:YNN}
   \mathbb{E}[Y_iD_{i(i)}^{\lor}|Z_{i(i)}=1] =& \mathbb{E}[Y_i(0,0)D_{i(i)}^{\lor}(1,1)]\\
   =&\mathbb{E}[Y_i(0,0)|D_i(1,1)=0,D_{(i)}(1,1)=0]\mathbb{P}[D_i(1,1)=0,D_{(i)}(1,1)=0]. \notag
\end{align}
where we first use \Cref{ass:iv}.1 and \ref{ass:iv}.2, and the same arguments as for the derivation of $\mathbb{E}[D_{i(i)}^{\lor}|Z_{i(i)}=1]$.

\subsection{Construction bounds $\tau_D(0)$}
Construct the lower bound as 
\begin{align}
   L^{10}_{00} =&  -\frac{\Delta^{10}_{00}\mathbb{E}[Y_i  D_{(i)}^{\lor}|Z]}{\Delta^{10}_{00}\mathbb{E}[D_{i(i)}^{\lor}|Z]}+\frac{\mathbb{E}[Y_iD_{i(i)}^{\lor}|Z_{i(i)}=1]}{\mathbb{E}[D_{i(i)}^{\lor}|Z_{i(i)}=1]}\frac{\Delta^{10}_{00}\mathbb{E}[ D_{(i)}^{\lor}|Z]}{\Delta^{10}_{00}\mathbb{E}[D_{i(i)}^{\lor}|Z]}\\
   =&\mathbb{E}[Y_i(1,0)-Y_i(0,0)|D_{i}(1,0) > D_{i}(0,0),D_{(i)}(0,0)=0]-\notag\\
   &(\mathbb{E}[Y_i(1,0)|D_{i}(1,0) > D_{i}(0,0),D_{(i)}(1,0)\neq D_{(i)}(0,0)=0]-\notag\\
   &\mathbb{E}[Y_i(0,0)|D_i(1,1)=0,D_{(i)}(1,1)=0])\times\notag\\
   &\frac{\mathbb{P}[D_{i}(1,0) > D_{i}(0,0),D_{(i)}(1,0)\neq D_{(i)}(0,0)=0]}{\mathbb{P}[D_{i}(1,0) > D_{i}(0,0), D_{(i)}(0,0)=0]}\\
   \leq&\mathbb{E}[Y_i(1,0)-Y_i(0,0)|D_{i}(1,0) > D_{i}(0,0),D_{(i)}(0,0)=0],
\end{align}
where we use that $\mathbb{E}[Y_i(1,0)|D_{i}(1,0) > D_{i}(0,0)]\geq\mathbb{E}[Y_i(0,0)|D_i(1,1)=0]$ according to the MTR assumption $\mathbb{E}[Y_i(1,0)|S_i,C_i] \geq \mathbb{E}[Y_i(0,0)|S_i,C_i]$ and the MTS assumption $\mathbb{E}[Y_i(0,0)|S_i,C_i]\geq\mathbb{E}[Y_i(0,0)|N_i]$. The first stages in the denominators exist if $-\Delta^{10}_{00}\mathbb{E}[D_{i(i)}^{\lor}|Z]=\mathbb{P}[\{S_i,C_i\}\times D_{(i)}(0,0)=0]>0$ and $\mathbb{E}[D_{i(i)}^{\lor}|Z_{i(i)}=1]=\mathbb{P}[N_i,N_{(i)}]>0$.

Construct the upper bound as 
\begin{align}
   U^{10}_{00} =&  -\frac{\Delta^{10}_{00}\mathbb{E}[Y_i  D_{(i)}^{\lor}|Z]}{\Delta^{10}_{00}\mathbb{E}[D_{i(i)}^{\lor}|Z]} +\frac{\Delta^{01}_{00}\mathbb{E}[Y_iD_iD_{(i)}^{\lor}|Z]}{\Delta^{01}_{00}\mathbb{E}[D_iD_{(i)}^{\lor}|Z]}\frac{\Delta^{10}_{00}\mathbb{E}[ D_{(i)}^{\lor}|Z]}{\Delta^{10}_{00}\mathbb{E}[D_{i(i)}^{\lor}|Z]}\\
      =&\mathbb{E}[Y_i(1,0)-Y_i(0,0)|D_{i}(1,0) > D_{i}(0,0),D_{(i)}(0,0)=0]+\notag\\
   &(\mathbb{E}[Y_i(1,0)|D_{i}(0,0)=1,D_{(i)}(0,1)\neq D_{(i)}(0,0)=0]-\notag\\
   &\mathbb{E}[Y_i(1,0)|D_{i}(1,0) > D_{i}(0,0),D_{(i)}(1,0)\neq D_{(i)}(0,0)=0])\times\notag\\
   &\frac{\mathbb{P}[D_{i}(1,0) > D_{i}(0,0),D_{(i)}(1,0)\neq D_{(i)}(0,0)=0]}{\mathbb{P}[D_{i}(1,0) > D_{i}(0,0), D_{(i)}(0,0)=0]}\\
   \geq&\mathbb{E}[Y_i(1,0)-Y_i(0,0)|D_{i}(1,0) > D_{i}(0,0),D_{(i)}(0,0)=0],
\end{align}
where we use that $\mathbb{E}[Y_i(1,0)|D_{i}(0,0)=1]\geq$  $\mathbb{E}[Y_i(1,0)|D_{i}(1,0) > D_{i}(0,0)]$ according to the MTS assumption $\mathbb{E}[Y_i(0,0)|A_i]\geq \mathbb{E}[Y_i(0,0)|S_i,C_i]$. The first stages in the denominators exist if $-\Delta^{10}_{00}\mathbb{E}[D_{i(i)}^{\lor}|Z]=\mathbb{P}[\{S_i,C_i\}\times D_{(i)}(0,0)=0]>0$ and $-\Delta^{01}_{00}\mathbb{E}[D_iD_{(i)}^{\lor}|Z]=\mathbb{P}[A_i\times \{ D_{(i)}(0,1) \neq D_{(i)}(0,0)=0 \}]>0$.

\subsection{First stages in bounds $\tau_D(1)$}
Derive $\Delta^{11}_{01}\mathbb{E}[ D_{(i)}^{\land}|Z]$, $\Delta^{11}_{01}\mathbb{E}[D_{i(i)}^{\land}|Z]$, $\Delta^{11}_{10}\mathbb{E}[(1-D_i)D_{(i)}^{\land}|Z]$, and $\mathbb{E}[D_{i(i)}^{\land}|Z_{i(i)}=0]$.
\begin{align}
    \Delta^{11}_{01}\mathbb{E}[ D_{(i)}^{\land}|Z] =& \mathbb{E}[D_{(i)}^{\land}(1,1)-D_{(i)}^{\land}(0,1)]\\
    =& \mathbb{P}[D_{i}(1,1) > D_{i}(0,1),  D_{(i)}(0,1) \neq D_{(i)}(1,1)=1],
\end{align}
where we use \Cref{ass:iv}.2, and subsequently that $D_{(i),j}(1,1) \geq D_{(i),j}(0,1)$ for all peers $j$ according to \Cref{ass:monopeer}, and if $D_{(i)}(1,1) \neq D_{(i)}(0,1)$, it follows from \Cref{ass:irrelevance} and \Cref{ass:iv}.3 that $D_{i}(1,1) > D_{i}(0,1)$.  
\begin{align}
    \Delta^{11}_{01}\mathbb{E}[D_{i(i)}^{\land}|Z]=&\mathbb{E}[D_{i(i)}^{\land}(1,1)-D_{i(i)}^{\land}(0,1)]\\
    =& \mathbb{P}[D_{i}(1,1) > D_{i}(0,1), D_{(i)}(1,1)= D_{(i)}(0,1)=1]+\notag\\
    & \mathbb{P}[D_{i}(1,1) > D_{i}(0,1), D_{(i)}(0,1)\neq D_{(i)}(1,1)= 1],
\end{align}
where we use \Cref{ass:iv}.2, and subsequently that $D_{(i),j}(1,1) \geq D_{(i),j}(0,1)$ for all peers $j$ according to \Cref{ass:monopeer}. If $D_{(i)}(1,1) = D_{(i)}(0,1)$, we require $D_{i}(1,0) \neq D_{i}(0,0)$ for a nonzero expression. If $D_{(i)}(1,1) \neq D_{(i)}(0,1)$, it follows from \Cref{ass:irrelevance} that $D_{i}(1,1) \neq D_{i}(0,1)$.  In both cases,  $D_{i}(1,1) > D_{i}(0,1)$ due to  \Cref{ass:iv}.3.
\begin{align}\label{eq:NC}
    \Delta^{11}_{10}\mathbb{E}[(1-D_i)D_{(i)}^{\land}|Z]=&\mathbb{E}[(1-D_i(1,1))D_{(i)}^{\land}(1,1)-(1-D_i(1,0))D_{(i)}^{\land}(1,0)]\\
    =& \mathbb{P}[D_i(1,1)=0,D_{(i)}(1,0) \neq D_{(i)}(1,1)=1].
\end{align}
where we use \Cref{ass:iv}.2, and subsequently that $D_i(1,1)\geq D_i(1,0)$ according to \Cref{ass:monopeer}. If $D_i(1,1) > D_i(1,0)$, it follows from \Cref{ass:irrelevance} that $D_{(i)}(1,1)\neq D_{(i)}(1,0)=0$ and the expression equals zero. If $D_i(1,1) = D_i(1,0)$, we require $ D_{(i)}(1,0) \neq D_{(i)}(1,1)=1$ for a nonzero expression. Using \Cref{ass:iv}.2, we have
\begin{align}\label{eq:AA}
    \mathbb{E}[D_{i(i)}^{\land}|Z_{i(i)}=0] = \mathbb{E}[D_{i(i)}^{\land}(0,0)]=\mathbb{P}[D_i(0,0)=1,D_{(i)}(0,0)=1].
\end{align}

\subsection{Reduced forms in bounds $\tau_D(1)$}
Derive expressions for $\Delta^{11}_{01}\mathbb{E}[Y_i  D_{(i)}^{\land}|Z]$, $\Delta^{11}_{10}\mathbb{E}[Y_i(1-D_i)D_{(i)}^{\land}|Z]$, and $\mathbb{E}[Y_iD_{i(i)}^{\land}|Z_{i(i)}=0]$.
\begin{align}
    \Delta^{11}_{01}\mathbb{E}[Y_i  D_{(i)}^{\land}|Z] =& \mathbb{E}[Y_i(D_i(1,1),1)D_{(i)}^{\land}(1,1)-Y_i(D_i(0,1),1)D_{(i)}^{\land}(0,1)]\\
    =& \mathbb{E}[Y_i(1,1)-Y_i(0,1)|D_{i}(1,1) > D_{i}(0,1), D_{(i)}(1,1)= D_{(i)}(0,1)=1]\notag\\
    &\times\mathbb{P}[D_{i}(1,1) > D_{i}(0,1), D_{(i)}(1,1)= D_{(i)}(0,1)=1]+\notag\\
    & \mathbb{E}[Y(1,1)|D_{i}(1,1) > D_{i}(0,1), D_{(i)}(0,1)\neq D_{(i)}(1,1)= 1]\notag\\
    &\times\mathbb{P}[D_{i}(1,1) > D_{i}(0,1), D_{(i)}(0,1)\neq D_{(i)}(1,1)= 1],
\end{align}
where we first use \Cref{ass:iv}.1 and \ref{ass:iv}.2, and use that $D_{(i)}^{\land}=0$ if there is a peer $j$ with $D_{(i),j}=0$. Second, we use the same arguments as for the derivation of $\Delta^{11}_{01}\mathbb{E}[  D_{i(i)}^{\land}|Z]$. Note that we can rewrite $\Delta^{11}_{01}\mathbb{E}[Y_i  D_{(i)}^{\land}|Z]$ to
\begin{align}
    \Delta^{11}_{01}\mathbb{E}[Y_i  D_{(i)}^{\land}|Z] =& \mathbb{E}[Y_i(1,1)-Y_i(0,1)|D_{i}(1,1) > D_{i}(0,1), D_{(i)}(1,1)=1]\notag\\
    &\times\mathbb{P}[D_{i}(1,1) > D_{i}(0,1), D_{(i)}(1,1)=1]+\notag\\
    & \mathbb{E}[Y(0,1)|D_{i}(1,1) > D_{i}(0,1), D_{(i)}(0,1)\neq D_{(i)}(1,1)= 1]\notag\\
    &\times\mathbb{P}[D_{i}(1,1) > D_{i}(0,1), D_{(i)}(0,1)\neq D_{(i)}(1,1)= 1].
\end{align}
\begin{align}\label{eq:YNC}
    \Delta^{11}_{10}\mathbb{E}[Y_i(1-D_i)D_{(i)}^{\land}|Z] =& \mathbb{E}[Y_i(0,1)\left( (1-D_i(1,1))D_{(i)}^\land(1,1)- (1-D_i(1,0))D_{(i)}^\land(1,0)\right)]\notag\\
    =&\mathbb{E}[Y_i(0,1)|D_i(1,1)=0,D_{(i)}(1,0) \neq D_{(i)}(1,1)=1]\notag\\
    &\times\mathbb{P}[D_i(1,1)=0,D_{(i)}(1,0) \neq D_{(i)}(1,1)=1],
\end{align}
where we first use \Cref{ass:iv}.1 and \ref{ass:iv}.2, and use that the expression is zero if $D_i=1$. Second, we use the same arguments as in the derivation of $\Delta^{11}_{10}\mathbb{E}[(1-D_i)D_{(i)}^{\land}|Z]$.
\begin{align}\label{eq:YAA}
    \mathbb{E}[Y_iD_{i(i)}^{\land}|Z_{i(i)}=0]=&\mathbb{E}[Y_i(1,1)|D_i(0,0)=1,D_{(i)}(0,0)=1]\notag\\
    &\times\mathbb{P}[D_i(0,0)=1,D_{(i)}(0,0)=1],
\end{align}
where we first use \Cref{ass:iv}.1 and \ref{ass:iv}.2, and the same arguments as in the derivation of $\mathbb{E}[D_{i(i)}^{\land}|Z_{i(i)}=0]$.

\subsection{Construction bounds $\tau_D(1)$}
Construct the lower bound as
\begin{align}
    L^{11}_{01} =&  \frac{\Delta^{11}_{01}\mathbb{E}[Y_i  D_{(i)}^{\land}|Z]}{\Delta^{11}_{01}\mathbb{E}[D_{i(i)}^{\land}|Z]}-\frac{\mathbb{E}[Y_iD_{i(i)}^{\land}|Z_{i(i)}=0]}{\mathbb{E}[D_{i(i)}^{\land}|Z_{i(i)}=0]}\frac{\Delta^{11}_{01}\mathbb{E}[ D_{(i)}^{\land}|Z]}{\Delta^{11}_{01}\mathbb{E}[D_{i(i)}^{\land}|Z]}\\
    =& \mathbb{E}[Y_i(1,1)-Y_i(0,1)|D_{i}(1,1) > D_{i}(0,1), D_{(i)}(1,1)=1]-\notag\\
    & (\mathbb{E}[Y(1,1)|D_{i}(0,0) =1, D_{(i)}(0,0)=1]-\notag\\
    &\mathbb{E}[Y(0,1)|D_{i}(1,1) > D_{i}(0,1), D_{(i)}(0,1)\neq D_{(i)}(1,1)= 1])\times\notag\\
    &\frac{\mathbb{P}[D_{i}(1,1) > D_{i}(0,1), D_{(i)}(0,1)\neq D_{(i)}(1,1)= 1]}{\mathbb{P}[D_{i}(1,1) > D_{i}(0,1), D_{(i)}(1,1)=1]},
\end{align}
where we use that $\mathbb{E}[Y(1,1)|D_{i}(0,0) =1]\geq\mathbb{E}[Y(0,1)|D_{i}(1,1) > D_{i}(0,1)]$ according to the MTR assumption $\mathbb{E}[Y_i(1,1)|A_i] \geq \mathbb{E}[Y_i(0,1)|A_i]$ and the MTS assumption $\mathbb{E}[Y(0,1)|A_i]\geq$  $\mathbb{E}[Y(0,1)|C_i,G_i]$. The first stages in the denominators exist if $\Delta^{11}_{01}\mathbb{E}[D_{i(i)}^{\land}|Z]=\mathbb{P}[\{C_i,G_i\}\times\{D_{(i)}(1,1)=1\}]>0$ and $\mathbb{E}[D_{i(i)}^{\land}|Z_{i(i)}=0]=\mathbb{P}[A_i,A_{(i)}]>0$.

Construct the upper bound as 
\begin{align}
    U^{11}_{01} =&\frac{\Delta^{11}_{01}\mathbb{E}[Y_i  D_{(i)}^{\land}|Z]}{\Delta^{11}_{01}\mathbb{E}[D_{i(i)}^{\land}|Z]}-\frac{\Delta^{11}_{10}\mathbb{E}[Y_i(1-D_i)D_{(i)}^{\land}|Z]}{\Delta^{11}_{10}\mathbb{E}[(1-D_i)D_{(i)}^{\land}|Z]}\frac{\Delta^{11}_{01}\mathbb{E}[ D_{(i)}^{\land}|Z]}{\Delta^{11}_{01}\mathbb{E}[D_{i(i)}^{\land}|Z]}\\
    =& \mathbb{E}[Y_i(1,1)-Y_i(0,1)|D_{i}(1,1) > D_{i}(0,1), D_{(i)}(1,1)=1]+\notag\\
    & (\mathbb{E}[Y(0,1)|D_{i}(1,1) > D_{i}(0,1), D_{(i)}(0,1)\neq D_{(i)}(1,1)= 1]-\notag\\
    &\mathbb{E}[Y_i(0,1)|D_i(1,1)=0,D_{(i)}(1,0) \neq D_{(i)}(1,1)=1])\times\notag\\
    &\frac{\mathbb{P}[D_{i}(1,1) > D_{i}(0,1), D_{(i)}(0,1)\neq D_{(i)}(1,1)= 1]}{\mathbb{P}[D_{i}(1,1) > D_{i}(0,1), D_{(i)}(1,1)=1]},
\end{align}
where we use that $\mathbb{E}[Y(0,1)|D_{i}(1,1) > D_{i}(0,1)]\geq\mathbb{E}[Y(0,1)|D_{i}(1,1) =0]$ according to the MTS assumption $\mathbb{E}[Y_i(0,1)|C_i,G_i]\geq \mathbb{E}[Y_i(0,1)|N_i]$. The first stages in the denominators exist if $\Delta^{11}_{01}\mathbb{E}[D_{i(i)}^{\land}|Z]=\mathbb{P}[\{C_i,G_i\}\times\{D_{(i)}(1,1)=1\}]>0$ and $\Delta^{11}_{10}\mathbb{E}[(1-D_i)D_{(i)}^{\land}|Z]=\mathbb{P}[N_i\times\{D_{(i)}(1,0)\neq D_{(i)}(1,1)=1\}]>0$.

\section{Proof \Cref{lemma:spillover}}
\subsection{First stages in bounds $\tau_S(0)$}
Derive expressions for $\Delta^{01}_{00}\mathbb{E}[D_i|Z]$, $\Delta^{01}_{00}\mathbb{E}[D_{i(i)}^{\lor}|Z]$, and use \eqref{eq:NC} and \eqref{eq:AA}.
\begin{align}\label{eq:fstauS0}
    \Delta^{01}_{00}\mathbb{E}[D_i|Z] &= \mathbb{E}[D_i(0,1)-D_i(0,0)] \\
    &= \mathbb{P}[D_i(0,1)>D_i(0,0),D_{(i)}(0,1) > D_{(i)}(0,0)],
\end{align}
where we use \Cref{ass:iv}.2 and \ref{ass:monopeer}, and subsequently \Cref{ass:irrelevance} and \ref{ass:iv}.3.
\begin{align}
    \Delta^{01}_{00}\mathbb{E}[D_{i(i)}^{\lor}|Z] =&  \mathbb{E}[D_{i(i)}^{\lor}(0,1)-D_{i(i)}^{\lor}(0,0)] \\
    =& -\mathbb{P}[D_i(0,1) = D_i(0,0)=0, D_{(i)}(0,1) > D_{(i)}(0,0)] \notag\\
     & -\mathbb{P}[D_i(0,1) > D_i(0,0), D_{(i)}(0,1) > D_{(i)}(0,0)],
\end{align}
where we use \Cref{ass:iv}.2, and subsequently that $D_i(0,1) \geq D_i(0,0)$ according to \ref{ass:monopeer}. If $D_i(0,1) > D_i(0,0)$, we require $D_{(i)}(0,1) > D_{(i)}(0,0)$ according to \Cref{ass:irrelevance} and \ref{ass:iv}.3. If $D_i(0,1) =D_i(0,0)=0$, we require $D_{(i)}(0,1) > D_{(i)}(0,0)$ for a nonzero expression.

\subsection{Reduced forms in bounds $\tau_S(0)$}
Derive an expression for $\Delta^{01}_{00}\mathbb{E}[Y_i(1-D_i)|Z]$, and use \eqref{eq:YNC} and \eqref{eq:YAA}.
\begin{align}
    \Delta^{01}_{00}\mathbb{E}[Y_i(1-D_i)|Z] =& \mathbb{E}[Y_i(0,D_{(i)}(0,1))(1-D_i(0,1))-Y_i(0,D_{(i)}(0,0))(1-D_i(0,0))]\notag\\
    =& \mathbb{E}[Y_i(0,1)-Y_i(0,0)|D_i(0,1) = D_i(0,0)=0, D_{(i)}(0,1) > D_{(i)}(0,0)]\notag\\
    &\times\mathbb{P}[D_i(0,1) = D_i(0,0)=0, D_{(i)}(0,1) > D_{(i)}(0,0)]\notag\\
     & -\mathbb{E}[Y_i(0,0)|D_i(0,1) > D_i(0,0), D_{(i)}(0,1) > D_{(i)}(0,0)]\notag\\
     &\times\mathbb{P}[D_i(0,1) > D_i(0,0), D_{(i)}(0,1) > D_{(i)}(0,0)],
\end{align}
where we first use \Cref{ass:iv}.1 and \ref{ass:iv}.2, and second use the same arguments as in the derivation of $\Delta^{01}_{00}\mathbb{E}[D_i|Z]$. We rewrite $ \Delta^{01}_{00}\mathbb{E}[Y_i(1-D_i)|Z]$ to
\begin{align}
    \Delta^{01}_{00}\mathbb{E}[Y_i(1-D_i)|Z] =&  \mathbb{E}[Y_i(0,1)-Y_i(0,0)|D_i(0,0) = 0, D_{(i)}(0,1) > D_{(i)}(0,0)]\notag\\
    &\times\mathbb{P}[D_i(0,0) = 0, D_{(i)}(0,1) > D_{(i)}(0,0)]\notag\\
     & -\mathbb{E}[Y_i(0,1)|D_i(0,1) > D_i(0,0), D_{(i)}(0,1) > D_{(i)}(0,0)]\notag\\
     &\times\mathbb{P}[D_i(0,1) > D_i(0,0), D_{(i)}(0,1) > D_{(i)}(0,0)].
\end{align}

\subsection{Construction bounds $\tau_S(0)$}
Construct the lower bound as
\begin{align}
    L^{01}_{00} =& -\frac{\Delta^{01}_{00}\mathbb{E}[Y_i(1-D_i)|Z]}{\Delta^{01}_{00}\mathbb{E}[D_{i(i)}^{\lor}|Z]}-\frac{\Delta^{11}_{10}\mathbb{E}[Y_i(1-D_i)D_{(i)}|Z]}{\Delta^{11}_{10}\mathbb{E}[(1-D_i)D_{(i)}|Z]}\frac{\Delta^{01}_{00}\mathbb{E}[D_{i}|Z]}{\Delta^{01}_{00}\mathbb{E}[D_{i(i)}^{\lor}|Z]}\\
      =& \mathbb{E}[Y_i(0,1)-Y_i(0,0)|D_i(0,0) = 0, D_{(i)}(0,1) > D_{(i)}(0,0)]-\notag\\
     & (\mathbb{E}[Y_i(0,1)|D_i(0,1) > D_i(0,0), D_{(i)}(0,1) > D_{(i)}(0,0)]-\notag\\
     &\mathbb{E}[Y_i(0,1)|D_i(1,1)=D_i(1,0)=0,D_{(i)}(1,1)>D_{(i)}(1,0)])\notag\\
     &\times\frac{\mathbb{P}[D_i(0,1) > D_i(0,0), D_{(i)}(0,1) > D_{(i)}(0,0)]}{\mathbb{P}[D_i(0,0) = 0, D_{(i)}(0,1) > D_{(i)}(0,0)]},
\end{align}
where we use that $\mathbb{E}[Y_i(0,1)|D_i(0,1) > D_i(0,0)]\geq\mathbb{E}[Y_i(0,1)|D_i(1,1)=D_i(1,0)=0]$ according to the MTS assumption $\mathbb{E}[Y_i(0,1)|S_i,P_i]\geq\mathbb{E}[Y_i(0,1)|N_i]$. The first stages in the denominators exist if $-\Delta^{01}_{00}\mathbb{E}[D_{i(i)}^{\lor}|Z]=\mathbb{P}[\{D_i(0,0)=0\}\times \{S_{(i)},C_{(i)}\}]>0$ and $\Delta^{11}_{10}\mathbb{E}[(1-D_i)D_{(i)}|Z]=\mathbb{P}[N_iC_{(i)}]>0$.

Construct the upper bound as
\begin{align}
    U^{01}_{00} =& -\frac{\Delta^{01}_{00}\mathbb{E}[Y_i(1-D_i)|Z]}{\Delta^{01}_{00}\mathbb{E}[D_{i(i)}^{\lor}|Z]}-\frac{\mathbb{E}[Y_iD_{i(i)}^{\land}|Z_{i(i)}=0]}{\mathbb{E}[D_{i(i)}^{\land}|Z_{i(i)}=0]}\frac{\Delta^{01}_{00}\mathbb{E}[D_{i}|Z]}{\Delta^{01}_{00}\mathbb{E}[D_{i(i)}^{\lor}|Z]}\\
      =& \mathbb{E}[Y_i(0,1)-Y_i(0,0)|D_i(0,0) = 0, D_{(i)}(0,1) > D_{(i)}(0,0)]+\notag\\
     & (\mathbb{E}[Y_i(1,1)|D_i(0,0)=1, D_{(i)}(0,0)=1]-\notag\\
     &\mathbb{E}[Y_i(0,1)|D_i(0,1) > D_i(0,0), D_{(i)}(0,1) > D_{(i)}(0,0)])\notag\\
     &\times\frac{\mathbb{P}[D_i(0,1) > D_i(0,0), D_{(i)}(0,1) > D_{(i)}(0,0)]}{\mathbb{P}[D_i(0,0) = 0, D_{(i)}(0,1) > D_{(i)}(0,0)]},
\end{align}
where we use that $\mathbb{E}[Y_i(1,1)|D_i(0,0)=1] \geq \mathbb{E}[Y_i(0,1)|D_i(0,1) > D_i(0,0)]$ according to the MTR assumption $\mathbb{E}[Y_i(1,1)|A_i] \geq \mathbb{E}[Y_i(0,1)|A_i]$ and the MTS assumption $\mathbb{E}[Y_i(0,1)|A_i] \geq \mathbb{E}[Y_i(0,1)|S_i,P_i]$. The first stages in the denominators exist if $-\Delta^{01}_{00}\mathbb{E}[D_{i(i)}^{\lor}|Z]=\mathbb{P}[\{D_i(0,0)=0\}\times \{S_{(i)},C_{(i)}\}]>0$ and $\mathbb{E}[D_{i(i)}^{\land}|Z_{i(i)}=0]=\mathbb{P}[A_iA_{(i)}]>0$.

\subsection{First stages in bounds $\tau_S(1)$}
Derive expressions for $\Delta^{11}_{10} \mathbb{E}[D_i|Z]$ and $\Delta^{11}_{10} \mathbb{E}[D_{i(i)}^\land|Z]$, and use \eqref{eq:AC} and \eqref{eq:NN}. 
\begin{align}
    \Delta^{11}_{10} \mathbb{E}[D_i|Z]=& \mathbb{E}[D_i(1,1)-D_i(1,0)]\\
    =& \mathbb{P}[D_i(1,1) > D_i(1,0),D_{(i)}(1,1) > D_{(i)}(1,0)],
\end{align}
where we use \Cref{ass:iv}.2 and \ref{ass:monopeer}, and subsequently \Cref{ass:irrelevance} and \ref{ass:iv}.3.
\begin{align}
    \Delta^{11}_{10} \mathbb{E}[D_{i(i)}^\land|Z] =& \mathbb{E}[D_{i(i)}^\land(1,1)-D_{i(i)}^\land(1,0)]\\
    =& \mathbb{P}[D_i(1,1)=D_i(1,0)=1,D_{(i)}(1,1)>D_{(i)}(1,0)]+\notag\\
     & \mathbb{P}[D_i(1,1)>D_i(1,0),D_{(i)}(1,1)>D_{(i)}(1,0)],
\end{align}
where we use \Cref{ass:iv}.2 and subsequently that $D_i(1,1)\geq D_i(1,0)$ according to \Cref{ass:monopeer}. If $D_i(1,1)=D_i(1,0)=1$, we require $D_{(i)}(1,1)>D_{(i)}(1,0)$ for a nonzero expression. If $D_i(1,1)>D_i(1,0)$, we require $D_{(i)}(1,1)>D_{(i)}(1,0)$ according to \Cref{ass:irrelevance} and \ref{ass:iv}.3.

\subsection{Reduced forms in bounds $\tau_S(1)$}
Derive an expression for $\Delta^{11}_{10} \mathbb{E}[Y_iD_i|Z]$, and use \eqref{eq:YAC} and \eqref{eq:YNN}.
\begin{align}
    \Delta^{11}_{10} \mathbb{E}[Y_iD_i|Z] =& \mathbb{E}[Y_i(1,D_{(i)}(1,1))D_i(1,1)-Y_i(1,D_{(i)}(1,0))D_i(1,0)]\\
    =&\mathbb{E}[Y_i(1,1)-Y_i(1,0)|D_i(1,1)=D_i(1,0)=1,D_{(i)}(1,1)>D_{(i)}(1,0)]\notag\\
    &\times \mathbb{P}[D_i(1,1)=D_i(1,0)=1,D_{(i)}(1,1)>D_{(i)}(1,0)]+\notag\\
    &\mathbb{E}[Y_i(1,1)|D_i(1,1)>D_i(1,0),D_{(i)}(1,1)>D_{(i)}(1,0)]\notag\\
    &\times \mathbb{P}[D_i(1,1)>D_i(1,0),D_{(i)}(1,1)>D_{(i)}(1,0)],
\end{align}
where we first use \Cref{ass:iv}.1 and \ref{ass:iv}.2, and subsequently use the same arguments as in the derivation of $\Delta^{11}_{10} \mathbb{E}[D_i|Z]$. We can rewrite $\Delta^{11}_{10} \mathbb{E}[Y_iD_i|Z]$ to
\begin{align}
    \Delta^{11}_{10} \mathbb{E}[Y_iD_i|Z] 
    =&\mathbb{E}[Y_i(1,1)-Y_i(1,0)|D_i(1,1)=1,D_{(i)}(1,1)>D_{(i)}(1,0)]\notag\\
    &\times \mathbb{P}[D_i(1,1)=1,D_{(i)}(1,1)>D_{(i)}(1,0)]+\notag\\
    &\mathbb{E}[Y_i(1,0)|D_i(1,1)>D_i(1,0),D_{(i)}(1,1)>D_{(i)}(1,0)]\notag\\
    &\times \mathbb{P}[D_i(1,1)>D_i(1,0),D_{(i)}(1,1)>D_{(i)}(1,0)].
\end{align}

\subsection{Construction bounds $\tau_S(1)$}
Construct the lower bound as
\begin{align}
    L^{11}_{10}=& \frac{\Delta^{11}_{10} \mathbb{E}[Y_iD_i|Z] }{\Delta^{11}_{10} \mathbb{E}[D_{i(i)}^\land|Z]}-\frac{\Delta^{01}_{00}\mathbb{E}[Y_iD_iD_{(i)}^{\lor}|Z]}{\Delta^{01}_{00}\mathbb{E}[D_iD_{(i)}^{\lor}|Z]}\frac{\Delta^{11}_{10} \mathbb{E}[D_{i}|Z]}{\Delta^{11}_{10} \mathbb{E}[D_{i(i)}^\land|Z]}\\
    =& \mathbb{E}[Y_i(1,1)-Y_i(1,0)|D_i(1,1)=1,D_{(i)}(1,1)>D_{(i)}(1,0)]-\notag\\
    &(\mathbb{E}[Y_i(1,0)|D_i(0,1)=D_i(0,0)=1,D_{(i)}(0,1)>D_{(i)}(0,0)]-\notag\\
    &\mathbb{E}[Y_i(1,0)|D_i(1,1)>D_i(1,0),D_{(i)}(1,1)>D_{(i)}(1,0)])\notag\\
    &\times \frac{\mathbb{P}[D_i(1,1)>D_i(1,0),D_{(i)}(1,1)>D_{(i)}(1,0)]}{\mathbb{P}[D_i(1,1)=1,D_{(i)}(1,1)>D_{(i)}(1,0)]},
\end{align}
where we use that $\mathbb{E}[Y_i(1,0)|D_i(0,0)=1]\geq \mathbb{E}[Y_i(1,0)|D_i(1,1)>D_i(1,0)]$ according to the MTS assumption $\mathbb{E}[Y_i(1,0)|A_i]\geq \mathbb{E}[Y_i(1,0)|P_i,G_i]$. The first stages in the denominators exist if $\Delta^{11}_{10} \mathbb{E}[D_{i(i)}^\land|Z]=\mathbb{P}[\{D_i(1,1)>D_i(1,0)\}\times\{C_i,G_i\}]$ and $-\Delta^{01}_{00}\mathbb{E}[D_iD_{(i)}^{\lor}|Z]=\mathbb{P}[A_i\times \{ D_{(i)}(0,1) \neq D_{(i)}(0,0)=0 \}]>0$.

Construct the upper bound as
\begin{align}
    U=& \frac{\Delta^{11}_{10} \mathbb{E}[Y_iD_i|Z] }{\Delta^{11}_{10} \mathbb{E}[D_{i(i)}^\land|Z]}-\frac{\mathbb{E}[Y_iD_{i(i)}^{\lor}|Z_{i(i)}=1]}{\mathbb{E}[D_{i(i)}^{\lor}|Z_{i(i)}=1]}\frac{\Delta^{11}_{10} \mathbb{E}[D_{i}|Z]}{\Delta^{11}_{10} \mathbb{E}[D_{i(i)}^\land|Z]}\\
    =& \mathbb{E}[Y_i(1,1)-Y_i(1,0)|D_i(1,1)=1,D_{(i)}(1,1)>D_{(i)}(1,0)]+\notag\\
    &(\mathbb{E}[Y_i(1,0)|D_i(1,1)>D_i(1,0),D_{(i)}(1,1)>D_{(i)}(1,0)]-\notag\\
    &\mathbb{E}[Y_i(0,0)|D_i(1,1)=0,D_{(i)}(1,1)=0])\notag\\
    &\times \frac{\mathbb{P}[D_i(1,1)>D_i(1,0),D_{(i)}(1,1)>D_{(i)}(1,0)]}{\mathbb{P}[D_i(1,1)=1,D_{(i)}(1,1)>D_{(i)}(1,0)]},
\end{align}
where we use that $\mathbb{E}[Y_i(1,0)|D_i(1,1)>D_i(1,0)]\geq \mathbb{E}[Y_i(0,0)|D_i(1,1)=0]$ according to the MTR assumption $\mathbb{E}[Y_i(1,0)|P_i,G_i] \geq \mathbb{E}[Y_i(0,0)|P_i,G_i]$ and the MTS assumption $\mathbb{E}[Y_i(0,0)|P_i,G_i] \geq \mathbb{E}[Y_i(0,0)|N_i]$. The first stages in the denominators exist if $\Delta^{11}_{10} \mathbb{E}[D_{i(i)}^\land|Z]=\mathbb{P}[\{D_i(1,1)>D_i(1,0)\}\times\{C_i,G_i\}]$ and $\mathbb{E}[D_{i(i)}^{\lor}|Z_{i(i)}=1]=\mathbb{P}[N_iN_{(i)}]>0$.

\section{Proof \Cref{theorem:spillover}}
This proof is similar to the proof of \Cref{lemma:spillover}, but here we have more than one peer and one-sided noncompliance. 

\subsection{Identification $\tau_S(0)$}
\begin{align}
    \Delta^{01}_{00}\mathbb{E}[D_{i(i)}^{\lor}|Z] =& \mathbb{E}[D_{i(i)}^{\lor}(0,1)-D_{i(i)}^{\lor}(0,0)]\\
    =& -\mathbb{P}[D_i(0,1)=D_i(0,0)=0,D_{(i)}(0,1)\neq D_{(i)}(0,0)=0],
\end{align}
where we use \Cref{ass:iv}.2, and subsequently that $D_i(0,1)=D_i(0,0)=0$ according to \Cref{ass:osnc}, and hence we require $D_{(i)}(0,1)\neq D_{(i)}(0,0)$ for a nonzero expression, with $D_{(i)}(0,0)=0$ according to \Cref{ass:osnc}. 

\begin{align}
    \Delta^{01}_{00}\mathbb{E}[Y_i(1-D_i)|Z] =& \mathbb{E}[Y_i(0,D_{(i)}(0,1))(1-D_i(0,1))-Y_i(0,D_{(i)}(0,0))(1-D_i(0,0))]\notag\\
    =& \mathbb{E}[Y_i(0,D_{(i)}(0,1))-Y_i(0,0)|D_i(0,1)=D_i(0,0)=D_{(i)}(0,0)=0]\notag\\
    &\times \mathbb{P}[D_{(i)}(0,1) \neq D_i(0,1)=D_i(0,0)=D_{(i)}(0,0)=0],
\end{align}
where we first use \Cref{ass:iv}.1 and \ref{ass:iv}.2, notice that $D_{(i)}(0,0)=0$ according to \Cref{ass:osnc}, and subsequently use the same arguments as for the derivation of $\Delta^{01}_{00}\mathbb{E}[D_{i(i)}^{\lor}|Z]$. 

It follows that $\tau_S(0)=-\Delta^{01}_{00}\mathbb{E}[Y_i(1-D_i)|Z]/\Delta^{01}_{00}\mathbb{E}[D_{i(i)}^{\lor}|Z]$ if $\Delta^{01}_{00}\mathbb{E}[D_{i(i)}^{\lor}|Z] = \mathbb{P}[D_{(i)}(0,1)\neq1]>0$.

\subsection{Identification $\tau_S(1)$}
First derive the first stages $\Delta^{11}_{10}\mathbb{E}[D_i|Z]$ and $\Delta^{11}_{10}\mathbb{E}[D_i D_{(i)}^{\lor}|Z]$, and use \eqref{eq:NN}.
\begin{align}
    \Delta^{11}_{10}\mathbb{E}[D_i|Z] =& \mathbb{E}[D_i(1,1)-D_i(1,0)]\\
    =& \mathbb{P}[D_i(1,1)>D_i(1,0),D_{(i)}(1,1) \neq D_{(i)}(1,0)=0],
\end{align}
where we use \Cref{ass:iv}.2, and subsequently \Cref{ass:monopeer}, \ref{ass:irrelevance}, and \ref{ass:osnc}.
\begin{align}
    \Delta^{11}_{10}\mathbb{E}[D_i D_{(i)}^{\lor}|Z] =& \mathbb{E}[D_i(1,1)D_{(i)}^{\lor}(1,1)-D_i(1,0)D_{(i)}^{\lor}(1,0)]\\
    =& -\mathbb{P}[D_i(1,1)=D_i(1,0)=1,D_{(i)}(1,1) \neq D_{(i)}(1,0)=0],
\end{align}
where we use \Cref{ass:iv}.2, and that $D_i(1,1)>D_i(1,0)$ requires  $D_{(i)}(1,1) \neq 0$ according to \Cref{ass:irrelevance}, which means the expression equals zero. It holds that $D_{(i)}(1,0)=0$ under \Cref{ass:osnc}.

Second, derive the reduced forms $\Delta^{11}_{10}\mathbb{E}[Y_iD_i|Z]$ and $\Delta^{11}_{10}\mathbb{E}[Y_iD_i D_{(i)}^{\lor}|Z]$, and use \eqref{eq:YNN}. 
\begin{align}
    \Delta^{11}_{10}\mathbb{E}[Y_iD_i|Z] =& \mathbb{E}[Y_i(1,D_{(i)}(1,1))D_i(1,1)-Y_i(1,D_{(i)}(1,0))D_i(1,0)]\\
    =& \mathbb{E}[Y_i(1,D_{(i)}(1,1))-Y_i(1,0)|D_i(1,1)=D_i(1,0)=1,D_{(i)}(1,0)=0]\notag\\
    &\times \mathbb{P}[D_i(1,1)=D_i(1,0)=1,D_{(i)}(1,1) \neq D_{(i)}(1,0)=0]+\notag\\
    & \mathbb{E}[Y_i(1,D_{(i)}(1,1))|D_i(1,1)>D_i(1,0),D_{(i)}(1,0)=0]\notag\\
    &\times \mathbb{P}[D_i(1,1)>D_i(1,0),D_{(i)}(1,1) \neq D_{(i)}(1,0)=0],
\end{align}
where we first use \Cref{ass:iv}.1 and \ref{ass:iv}.2, and subsequently that $D_i(1,1) \geq D_i(1,0) \geq 0 $ and $D_{(i)}(1,0)=0$ according to \Cref{ass:monopeer} and \ref{ass:osnc}. We rewrite $\Delta^{11}_{10}\mathbb{E}[Y_iD_i|Z]$ to
\begin{align}
    \Delta^{11}_{10}\mathbb{E}[Y_iD_i|Z]    =& \mathbb{E}[Y_i(1,D_{(i)}(1,1))-Y_i(1,0)|D_i(1,1)=1,D_{(i)}(1,0)=0]\\
    &\times \mathbb{P}[D_i(1,1)=1,D_{(i)}(1,1) \neq D_{(i)}(1,0)=0]+\notag\\
    & \mathbb{E}[Y_i(1,0)|D_i(1,1)>D_i(1,0), D_{(i)}(1,0)=0]\notag\\
    &\times \mathbb{P}[D_i(1,1)>D_i(1,0),D_{(i)}(1,1) \neq D_{(i)}(1,0)=0].
\end{align}
\begin{align}
    \Delta^{11}_{10}\mathbb{E}[Y_iD_i D_{(i)}^{\lor}|Z] =& \mathbb{E}[Y_i(1,0)(D_i(1,1)D_{(i)}^{\lor}(1,1)-D_i(1,0)D_{(i)}^{\lor}(1,0))]\\
    =& -\mathbb{E}[Y_i(1,0)|D_i(1,1)=D_i(1,0)=1,D_{(i)}(1,1) \neq D_{(i)}(1,0)=0]\notag\\
    &\times \mathbb{P}[D_i(1,1)=D_i(1,0)=1,D_{(i)}(1,1) \neq D_{(i)}(1,0)=0],
\end{align}
where we use \Cref{ass:iv}.1 and \ref{ass:iv}.2, and subsequently  use the same arguments as for the derivation of $\Delta^{11}_{10}\mathbb{E}[D_i D_{(i)}^{\lor}|Z]$.

Construct the lower bound as
\begin{align}
    L^{11}_{10} =& \frac{\Delta^{11}_{10}\mathbb{E}[Y_iD_i|Z]}{\Delta^{11}_{10}\mathbb{E}[D_i(1- D_{(i)}^{\lor})|Z]}-\frac{\Delta^{11}_{10}\mathbb{E}[Y_iD_i D_{(i)}^{\lor}|Z]}{\Delta^{11}_{10}\mathbb{E}[D_i D_{(i)}^{\lor}|Z]}\frac{\Delta^{11}_{10}\mathbb{E}[D_i|Z]}{\Delta^{11}_{10}\mathbb{E}[D_i(1- D_{(i)}^{\lor})|Z]}\\
    =& \mathbb{E}[Y_i(1,D_{(i)}(1,1))-Y_i(1,0)|D_i(1,1)=1,D_{(i)}(1,0)=0]-\notag\\
    &(\mathbb{E}[Y_i(1,0)|D_i(1,1)=D_i(1,0)=1,D_{(i)}(1,1) \neq D_{(i)}(1,0)=0]-\notag\\
    & \mathbb{E}[Y_i(1,0)|D_i(1,1)>D_i(1,0), D_{(i)}(1,0)=0])\times\notag\\
    &\frac{\mathbb{P}[D_i(1,1)>D_i(1,0),D_{(i)}(1,1) \neq D_{(i)}(1,0)=0]}{\mathbb{P}[D_i(1,1)=1,D_{(i)}(1,1) \neq D_{(i)}(1,0)=0]},
\end{align}
where we use that $\mathbb{E}[Y_i(1,0)|D_i(1,1)=D_i(1,0)=1] \geq \mathbb{E}[Y_i(1,0)|D_i(1,1)>D_i(1,0)]$ according to the MTS assumption $\mathbb{E}[Y_i(1,0)|C_i] \geq \mathbb{E}[Y_i(1,0)|G_i]$. The first stages in the denominators exist if $\Delta^{11}_{10}\mathbb{E}[D_i(1- D_{(i)}^{\lor})|Z]=\mathbb{P}[\{C_i,G_i\}\times \{D_i(1,1)\neq 0\}]>0$ and $\Delta^{11}_{10}\mathbb{E}[D_i D_{(i)}^{\lor}|Z]=\mathbb{P}[C_i\times \{D_i(1,1)\neq 0\}]>0$.

Construct the upper bound as
\begin{align}
    U^{11}_{10} =& \frac{\Delta^{11}_{10}\mathbb{E}[Y_iD_i|Z]}{\Delta^{11}_{10}\mathbb{E}[D_i(1- D_{(i)}^{\lor})|Z]}-\frac{\mathbb{E}[Y_iD_{i(i)}^{\lor}|Z_{i(i)}=1]}{\mathbb{E}[D_{i(i)}^{\lor}|Z_{i(i)}=1]}\frac{\Delta^{11}_{10}\mathbb{E}[D_i|Z]}{\Delta^{11}_{10}\mathbb{E}[D_i(1- D_{(i)}^{\lor})|Z]}\\
    =& \mathbb{E}[Y_i(1,D_{(i)}(1,1))-Y_i(1,0)|D_i(1,1)=1,D_{(i)}(1,0)=0]+\notag\\
    &(\mathbb{E}[Y_i(1,0)|D_i(1,1)>D_i(1,0), D_{(i)}(1,0)=0]-\notag\\
    & \mathbb{E}[Y_i(0,0)|D_i(1,1)=0,D_{(i)}(1,1)=0])\times\notag\\
    &\frac{\mathbb{P}[D_i(1,1)>D_i(1,0),D_{(i)}(1,1) \neq D_{(i)}(1,0)=0]}{\mathbb{P}[D_i(1,1)=1,D_{(i)}(1,1) \neq D_{(i)}(1,0)=0]},
\end{align}
where we use that $\mathbb{E}[Y_i(1,0)|D_i(1,1)>D_i(1,0)]\geq \mathbb{E}[Y_i(0,0)|D_i(1,1)=0]$ according to the MTR assumption $\mathbb{E}[Y_i(1,0)|G_i]\geq \mathbb{E}[Y_i(0,0)|G_i]$ and the MTS assumption $\mathbb{E}[Y_i(0,0)|G_i]\geq \mathbb{E}[Y_i(0,0)|N_i]$. The first stages in the denominators exist if $\Delta^{11}_{10}\mathbb{E}[D_i(1- D_{(i)}^{\lor})|Z]=\mathbb{P}[\{C_i,G_i\}\times \{D_i(1,1)\neq 0\}]>0$ and $\mathbb{E}[D_{i(i)}^{\lor}|Z_{i(i)}=1]=\mathbb{P}[N_iN_{(i)}]>0$.

\section{Absence particular compliance types combinations}
In the absence of particular compliance types, denominators in the bounds equal zero, and these bounds do not exist. The bounds in this paper make use of the potential outcomes of five different combinations of compliance types. This appendix discusses alternative potential outcomes that can be used in the bounds in case these compliance types are absent. 

\subsection{Alternative potential outcomes}
Denote by $Y_{\min}$ and $Y_{\max}$ the smallest and largest possible outcome values. Second, 
\begin{align}
    \mathbb{E}[D_i D_{(i)}^\lor|Z_i=0,Z_{(i)}=1] =& \mathbb{E}[D_i(0,1) D_{(i)}^\lor(0,1)]
    = \mathbb{P}[D_i(0,1)=1, D_{(i)}(0,1)=0]\\
    =& \mathbb{P}[D_i(1,1)=D_i(0,1)=1, D_{(i)}(0,1)=D_{(i)}(0,0)=0]\\
    =& \mathbb{P}[D_i(0,0)=1, D_{(i)}(1,1)=0] = \mathbb{P}[A_i,N_{(i)}],
\end{align}
where we use \Cref{ass:iv}.2 and \ref{ass:iv}.3, and use that according to \Cref{ass:irrelevance} $D_{(i)}(0,1)=D_{(i)}(1,1)$ if $D_{i}(0,1)=D_{i}(1,1)$, and $D_{i}(0,1)=D_{i}(0,0)$ if $D_{(i)}(0,1)=D_{(i)}(0,0)$. Using \Cref{ass:iv}.1 and the arguments above,
\begin{align}
    \mathbb{E}[Y_i D_i D_{(i)}^\lor|Z_i=0,Z_{(i)}=1] = \mathbb{E}[Y_i(1,0)|A_i,N_{(i)}]\mathbb{P}[A_i,N_{(i)}].
\end{align}
Third, we derive
\begin{align}
    \mathbb{E}[(1-D_i) D_{(i)}^\land|Z_i=1,Z_{(i)}=0] =& \mathbb{E}[(1-D_i(1,0)) D_{(i)}^\land(1,0)]=\mathbb{P}[D_i(1,0)=0,D_{(i)}(1,0)=1]\notag\\
    =& \mathbb{P}[D_i(1,0)=D_i(0,0)=0,D_{(i)}(1,1)=D_{(i)}(1,0)=1]\notag\\
    =& \mathbb{P}[D_i(1,1)=0,D_{(i)}(0,0)=1] =\mathbb{P}[N_i,A_{(i)}],
\end{align}
where we use \Cref{ass:iv}.2 and \ref{ass:iv}.3, and use that according to \Cref{ass:irrelevance} $D_{(i)}(1,0)=D_{(i)}(0,0)$ if $D_i(1,0)=D_i(0,0)$, and $D_{i}(1,1)=D_{i}(1,0)$ if $D_{(i)}(1,1)=D_{(i)}(1,0)$. Using \Cref{ass:iv}.1 and the arguments above,
\begin{align}
    \mathbb{E}[Y_i (1-D_i) D_{(i)}^\land|Z_i=1,Z_{(i)}=0] = \mathbb{E}[Y_i(0,1)|N_i,A_{(i)}]\mathbb{P}[N_i,A_{(i)}].
\end{align}

\subsection{Alternative bounds}
Consider $L^{10}_{00}$ in \Cref{lemma:direct} and \Cref{theorem:direct}, and $U_{11}^{10}$ in \Cref{lemma:spillover} and \Cref{theorem:spillover}.
These bounds rely on $\mathbb{E}[Y_iD_{i(i)}^\lor|Z_{i(i)}=1]/\mathbb{E}[D_{i(i)}^\lor|Z_{i(i)}=1]$. If $\mathbb{E}[D_{i(i)}^\lor|Z_{i(i)}=1]=\mathbb{P}[N_iN_{(i)}]=0$, this can be replaced by $Y_{\min}$.

Consider $U^{10}_{00}$ in \Cref{lemma:direct} and \Cref{theorem:direct}, and $L_{11}^{10}$ in \Cref{lemma:spillover}.
These bounds rely on $\Delta^{01}_{00}\mathbb{E}[Y_iD_iD_{(i)}^\lor|Z]/\Delta^{01}_{00}\mathbb{E}[D_iD_{(i)}^\lor|Z]$. If $\Delta^{01}_{00}\mathbb{E}[D_iD_{(i)}^\lor|Z]=\mathbb{P}[A_i\times \{D_{(i)}(0,1)\neq D_{(i)}(0,0)=0\}]=0$, this can be replaced by $Y_{\max}$ or by $\mathbb{E}[Y_i D_i D_{(i)}^\lor|Z_i=0,Z_{(i)}=1]/\mathbb{E}[D_i D_{(i)}^\lor|Z_i=0,Z_{(i)}=1]$ if $\mathbb{P}[A_i,N_{(i)}]$ exists.

Consider $L^{11}_{01}$ in \Cref{lemma:direct} and \Cref{theorem:direct}, and $U_{00}^{01}$ in \Cref{lemma:spillover}.
These bounds rely on $\mathbb{E}[Y_iD_{i(i)}^\land|Z_{i(i)}=0]/\mathbb{E}[D_{i(i)}^\land|Z_{i(i)}=0]$. If $\mathbb{E}[D_{i(i)}^\land|Z_{i(i)}=0]=\mathbb{P}[A_iA_{(i)}]=0$, this can be replaced by $Y_{\max}$.

Consider $U^{11}_{01}$ in \Cref{lemma:direct} and \Cref{theorem:direct}, and $L_{00}^{01}$ in \Cref{lemma:spillover}.
These bounds rely on $\Delta^{11}_{10}\mathbb{E}[Y_i(1-D_i)D_{(i)}^\land|Z]/\Delta^{11}_{10}\mathbb{E}[(1-D_i)D_{(i)}^\land|Z]$. If $\Delta^{11}_{10}\mathbb{E}[(1-D_i)D_{(i)}^\land|Z]=\mathbb{P}[N_i\times \{D_{(i)}(1,0)\neq D_{(i)}(1,1)=1\}]=0$, this can be replaced by $Y_{\min}$ or by $\mathbb{E}[Y_i (1-D_i) D_{(i)}^\land|Z_i=1,Z_{(i)}=0]/\mathbb{E}[(1-D_i) D_{(i)}^\land|Z_i=1,Z_{(i)}=0]$ if $\mathbb{P}[N_i,A_{(i)}]$ exists.

Consider $L^{11}_{01}$ in \Cref{theorem:spillover}, which relies on $\Delta^{11}_{10}\mathbb{E}[Y_iD_iD_{(i)}^\lor|Z]/\Delta^{11}_{10}\mathbb{E}[D_iD_{(i)}^\lor|Z]$. If $\Delta^{11}_{10}\mathbb{E}[D_iD_{(i)}^\lor|Z]=\mathbb{P}[C_i\times \{D_{(i)}(1,1)\neq 0\}]=0$, this can be replaced by $Y_{\max}$.

\subsection{One-sided noncompliance}
Under \Cref{ass:osnc}, always-takers, social compliers, and peer compliers are absent and the bounds in \Cref{lemma:direct}, \Cref{lemma:spillover}, and \Cref{theorem:direct} have to be adjusted.  

The lower bounds $L^{10}_{00}$ in \Cref{lemma:direct} and \Cref{theorem:direct} and $L^{11}_{10}$ in \Cref{lemma:spillover} can be adjusted with $y_{\max}$ as described above. 

It follows from \eqref{eq:fstauD0} that $\Delta^{10}_{00}\mathbb{E}[ D_{(i)}^{\lor}|Z]=0$ in the absence of social and peer compliers, and hence $L^{10}_{00}=U^{10}_{00}=\tau_D(0)$ in \Cref{lemma:direct} and \Cref{theorem:direct}. 

It follows from \eqref{eq:fstauS0} that $\Delta^{01}_{00}\mathbb{E}[D_i|Z]=0$ in the absence of social and peer compliers, and hence $L^{01}_{00}=U^{01}_{00}=\tau_S(0)$ in \Cref{lemma:spillover}.

\section{Necessary conditions irrelevance}
Consider the first inequality in \Cref{prop:irrelevance} with $d=0$:
\begin{align}
    \Delta^{10}_{00}\mathbb{E}[D_iD_{(i)}^\lor|Z]=& \mathbb{E}[D_i(1,0)D_{(i)}^\lor(1,0)-D_i(0,0)D_{(i)}^\lor(0,0)]\\
    =& \mathbb{P}[D_i(1,0)>D_i(0,0),D_{(i)}(1,0)=D_{(i)}(0,0)=0]-\notag\\
    & \mathbb{P}[D_i(1,0)=D_i(0,0)=1,D_{(i)}(1,0) \neq D_{(i)}(0,0)=0 ],
\end{align}
where we use \Cref{ass:iv}.2, and subsequently \Cref{ass:iv}.3 and \ref{ass:monopeer}. Note that under \Cref{ass:irrelevance}, $D_{(i)}(0,0) = D_{(i)}(1,0)$ if $D_i(1,0)=D_i(0,0)$, and hence $\Delta^{10}_{00}\mathbb{E}[D_iD_{(i)}^\lor|Z]\geq 0$.

Using the same arguments, we have for $d=0$ in \Cref{prop:irrelevance} that
\begin{align}
    \Delta^{10}_{00}\mathbb{E}[(1-D_i)D_{(i)}^\land|Z]=& \mathbb{P}[D_i(1,0)=D_i(0,0)=0,D_{(i)}(0,0) \neq D_{(i)}(1,0)=1]-\notag\\
    & \mathbb{P}[D_i(1,0)>D_i(0,0),D_{(i)}(1,0)=D_{(i)}(0,0) =1],\\
    \Delta^{01}_{00}\mathbb{E}[D_iD_{(i)}^\lor|Z]=&\mathbb{P}[D_i(0,1)>D_i(0,0),D_{(i)}(0,1)=D_{(i)}(0,0)=0]-\notag\\
    & \mathbb{P}[D_i(0,1)=D_i(0,0)=1,D_{(i)}(0,1) \neq D_{(i)}(0,0)=0],\\
    \Delta^{01}_{00}\mathbb{E}[(1-D_i)D_{(i)}^\land|Z]=& \mathbb{P}[D_i(0,1)=D_i(0,0)=0,D_{(i)}(0,0)\neq D_{(i)}(0,1)=1]-\notag\\
    &\mathbb{P}[D_i(0,1)>D_i(0,0)=0,D_{(i)}(0,1)=D_{(i)}(0,0)=1],
\end{align}
which satisfy the inequalities in \Cref{prop:irrelevance} under \Cref{ass:irrelevance}. Similarly, it follows for $d=1$.

\end{document}